\documentclass[fleqn,usenatbib]{mnras}

\usepackage{newtxtext,newtxmath}
\usepackage[T1]{fontenc}
\usepackage{ae,aecompl}

\usepackage{graphicx}
\usepackage{natbib}
\usepackage{amsmath}
\usepackage{lscape}
\usepackage{xcolor}
\usepackage{rotating}
\usepackage{booktabs}
\usepackage{caption}
\usepackage{ulem}
\usepackage{psfrag}
\usepackage{pstool}
\usepackage{multirow}
\usepackage{tikz}
\usepackage{mathtools}

\newsavebox{\astrutbox}
\sbox{\astrutbox}{\rule[-5pt]{0pt}{20pt}}

\def\<{\langle}
\def\>{\rangle}

\def\be{\begin{equation}}
\def\ee{\end{equation}}
\def\bea{\begin{eqnarray}}
\def\eea{\end{eqnarray}}

\providecommand{\av}[1]{\langle{#1}\rangle}

\DeclarePairedDelimiter\abs{\lvert}{\rvert}%

\def\aap{{\it  Astron.\ Astrophys}.}

\def\apss{{\it  Astrophys.\ Space Sci}.}

\def\araa{{\it  Ann.\ Rev.\ Astron.\ Astrophys}.}
\def\apj{{\it  Astrophys.\ J}.}
\def\apjs{{\it  Astrophys.\ J.\ Suppl}.}

\def\ajp{{\it  Aust.\ J.\ Phys}.}

\def\jpp{{\it  J.\ Plasma Phys}.}
\def\mnras{{\it  Mon.\ Not.\ R.\ Astron.\ Soc}.}

\def\nat{{\it  Nature} }

\def\prl{{\it  Phys.\ Rev.\ Lett}.}

\def\physscr{{\it  Physica Scripta}}

\def\spjetp{{\it  Sov.\ Phys.\ JETP} }
\def\sjpp{{\it  Sov.\ J.\ Plasma Phys.} }
\def\ssr{{\it  Space Sci.\ Rev}.}

\newcommand{\cma}{\color{magenta}}
\newcommand{\br}{\bf\cma}

\title[Pulsar radio emission mechanisms]{Pulsar radio emission mechanisms: a critique}

\author[D. B. Melrose, M. Z. Rafat and A. Mastrano]%
{D. B. Melrose$^1$, M. Z. Rafat$^{1}$ and A. Mastrano$^{1}$\\
$^1$SIfA, School of Physics, University of Sydney, Sydney, NSW 2006, Australia}

\date{Accepted XXX. Received YYY; in original form ZZZ}

\pubyear{2019}

\begin{document}
\label{firstpage}
\pagerange{\pageref{firstpage}--\pageref{lastpage}}
\maketitle

\begin{abstract}
We consider critically the three most widely favored pulsar radio emission mechanisms: coherent curvature emission (CCE), beam-driven relativistic plasma emission (RPE) and anomalous Doppler emission (ADE). We assume that the pulsar plasma is one dimensional (1D), streaming outward with a bulk Lorentz factor $\gamma_{\rm s} \gg \av{\gamma} -1 \gtrsim 1$, where $\av{\gamma}$ is the intrinsic spread in the rest frame of the plasma. We argue that the formation of beams in a multi-cloud model is ineffective in the intrinsically relativistic case for plausible parameters, because the overtaking takes too long. We argue that the default choice for the particle distribution in the rest frame is a J{\"u}ttner distribution and that relativistic streaming should be included by applying a Lorentz transformation to the rest-frame distribution, rather than the widely assumed relativistically streaming Gaussian distribution. We find that beam-driven wave growth is severely restricted by (a) the wave properties in pulsar plasma, (b) a separation condition between beam and background, and (c) the inhomogeneity of the plasma in the pulsar frame. The growth rate for the kinetic instability is much smaller and the bandwidth of the growing waves is much larger for a J{\"u}ttner distribution than for a relativistically streaming Gaussian distribution. No reactive instability occurs at all for a J{\"u}ttner distribution. We conclude that none of CCE, RPE and ADE in tenable as the generic pulsar radio emission mechanism for ``plausible'' assumptions about the pulsar plasma.
\end{abstract}


\section{Introduction}

After 50 years of research on radio pulsars, the mechanism by which pulsar radio emission is generated remains an enigma. Several different emission mechanisms were suggested in the 1970s and these have continued to attract both supporters and critics over the decades, but no consensus has been reached. We refer to the three mechanisms favored in the 1970s as coherent curvature emission (CCE), relativistic plasma emission (RPE) and anomalous Doppler emission (ADE), each of which is defined and discussed briefly below. For present purposes, we regard two other suggested mechanisms, linear acceleration emission (LAE) and free-electron maser emission (FEM), as variants of RPE.\footnote{As pointed out by \citet{Lyubarsky96}, LAE, FEM and induced scattering may all be interpreted as a second stage in RPE.} These early suggested mechanisms were based on two assumptions: first, that the emission occurs in polar-cap regions, defined by magnetic field lines that do not close within the light-cylinder radius and, second, that the ultimate source of the radiant energy is through ``primary'' particles accelerated from the surface of the star in the polar-cap regions, with ``secondary'' particles  generated by pair cascades populating the polar-cap regions with outflowing relativistic pairs. 

Our purpose in this paper is to discuss CCE, RPE and ADE critically, to determine whether any of them is viable as the generic pulsar radio emission mechanism. Our working hypothesis is that the similar features in all pulsar radio emission is indicative of a single generic radio emission mechanism operating in all pulsars. Such a generic mechanism should not be dependent on specialized assumptions nor should it be restricted to specifically favorable locations, but should be robust enough to account for essentially all pulsar radio emission. We retain the assumption that the emission source is in the polar-cap regions in outflowing relativistic pair plasma, so that we do not consider alternative mechanisms, such as radio emission generated by reconnection in the plasma sheet beyond the light cylinder \citep{2019ApJ...876L...6P,2019MNRAS.483.1731L}. We also argue against the assumption that the surface of the star is an important source of ``primary'' particles, in particular, excluding ``multiple-sparking'' models involving hot-spots on the stellar surface  \citep{RS75,FR_82,Beskin_82,Gil_S00}.

An important qualitative point concerning models based on charges drawn from the stellar surface is that they lead to a charge-separated electrosphere and not a polar-cp model. Charge-separated models were proposed in the mid 1970s \citep{R76,J76}. They were later developed in more detail \citep{KPM85a}, when the name electrosphere was coined \citep{KPM85b}. These models may be described as dome-disk models in which charges of one sign form domes above the poles and charges of the opposite sign form an equatorial disk. One criticism of early dome-disk models was that they apply only to the aligned case  and are unstable, but a contrary argument was that the dome-disk model is the state to which an aligned rotator relaxes \citep{SMT01}. Another criticism is that oblique dome-disk models are unstable to the diocotron instability. The diocotron instability \citep{PHB02} involves growing surface waves when two sheets of charge slip past one another; it may be regarded as an analog of the Kelvin-Helmholtz instability. However, detailed numerical modeling \citep{S04,Mi04} did not support this criticism, showing that oblique dome-disk models are robust. Here we assume that charges drawn from the stellar surface play no important role.  In more recent models, in which the intrinsic time-dependence is taken into account \citep[e.g.,][]{T10b,TA13}, the pair creation exhibits a limit cycle behavior that could be considered similar to what is assumed in a sparking model. We use the name ``multiple-beam'' to refer to any model in which pair cascades result in  localized transient clouds.  

The properties of the pulsar plasma in the source region of the radio emission play a central role in any discussion of the radio emission mechanism. Despite significant changes in our understanding of pulsar electrodynamics since the 1970s, the general properties of the pulsar plasma established at that time \citep[e.g.,][]{A79} have not changed greatly. It is widely accepted that the ``pulsar plasma'' is a relativistically outflowing, strongly magnetized, one-dimensional (1D),  electron-positron plasma, created by pair cascades, with a streaming  Lorentz factor $\gamma_{\rm s}\gg1$, and with a relativistic spread, $\av{\gamma} - 1 \gtrsim 1$, in its rest frame. Models that do not rely on primary particles from the stellar surface \citep{Letal05,B08,Lyubarsky09,T10b}, and Particle-in-Cell calculations \citep{TA13,PSC15,Cetal16,CB17,Ketal17,Ketal18,Betal18}, have not led to radically different models for the (time-averaged) properties of the bulk of the  pair plasma, compared with earlier models. Here we assume that ``plausible'' parameters for the pulsar plasma correspond to pairs streaming outward with a bulk Lorentz factor $\gamma_{\rm s}$ of order $10^2$--$10^3$, and with an intrinsic relativistic spread with $\av{\gamma} $ between a few and about 10 \citep{HA01,AE02}. Numerical models for pair cascades also imply the ratio, $\kappa$, between the number density of pairs and $\rho'_{\rm cor}/e$, where $\rho'_{\rm cor}$ is the corotation charge density. A plausible value is $\kappa=10^5$ \citep{TH15}.

In discussions of the radio emission mechanism, a variety of different assumptions have been made, either explicitly or implicitly, concerning the value of $\av{\gamma}$, including the assumption that the plasma is either cold, $\av{\gamma}\to1$, or nonrelativistic, $\av{\gamma}-1\ll1$, rather than intrinsically relativistic, $\av{\gamma}-1\gtrsim1$, in its rest frame. We assume that ``plausible'' properties for the pulsar plasma are $\gamma_{\rm s}\gg1$, $\av{\gamma}-1\gtrsim1$,  $\kappa=10^5$. We further quantify what we mean by ``plausible'' parameters in Section~\ref{sect:parameters}.

We define CCE, RPE and ADE and comment briefly on some of the arguments for and against each of them. 

\noindent
{\bf CCE}: Curvature emission by a single particle with Lorentz factor $\gamma\gg1$ in 1D motion along a curved magnetic  field line, with radius of curvature $R_c$, has a characteristic frequency $\omega=(c/R_c)\gamma^3$, with the frequency spectrum increasing $\propto\omega^{1/3}$ below this frequency, and falling off rapidly at higher frequencies. The basic assumption in early versions of CCE \citep{R69,K70,S71,RS75} is that the particles emit coherently at low frequencies, in the sense that $N$ particles radiate $N^2$ times the power per individual particle. There are qualitative properties of curvature emission that lead to it continuing to be the favored mechanism for the interpretation of observed features in the radio emission \citep[e.g.,][]{MGP00,GLM04,DRR07,MGM09,Mitra17}. However, the coherence mechanism for CCE has long been recognized as problematic (further discussion in Section~\ref{sect:CCE}). 

\noindent
{\bf RPE}: Plasma emission, for example in solar radio bursts, is a multi-stage emission process, and RPE is defined here as the relativistic counterpart of plasma emission \citep{M17a}. The first stage in plasma emission is an electron beam causing Langmuir waves to grow, and the other stages involve nonlinear processes (or the effect of inhomogeneities) partly converting the energy in the Langmuir waves into escaping radiation at the plasma frequency, $\omega_{\rm p}$, and its second harmonic.  Various versions of beam-driven RPE were suggested in the early literature on pulsar radio emission \citep{TK72,SC73,SC75,H76b,H76a,HR76,HR78,BB77,LM79,LMS79,LP82,APS83,ELM83,L92,A93,A95,W94}.  Resonant beam-driven growth occurs when the resonance condition, written here as $z=\beta$, applies with $\beta<\beta_{\rm b}$, where $z=\omega/k_\parallel c$ is the phase speed of the wave, $ \beta $ is the particle speed and  $\beta_{\rm b}$ is the beam speed. (We define a speed $\beta$ relative to the speed of light,  with $\gamma=(1-\beta^2)^{-1/2}$  the corresponding Lorentz factor and $u=\gamma\beta$ the 4-speed.) The assumption that the growing waves are ``Langmuir-like'' was questioned in some of the early literature, and led to the alternative suggestion that the beam generates Alfv\'en waves \citep{TK72,Letal82,MG99,L00}. Two other difficulties were recognized in early discussions of beam-driven  RPE. First, the growth rates for various suggested instabilities in the first stage are too slow to be effective \citep[e.g.,][]{BB77,ELM83,Letal86}, and it was suggested that this can be overcome by appealing to what was called a  multiple-sparking model in the older literature \citep{U87,UU88,AM98,U02,GMG02}, cf. Section~\ref{sect:multiple-sparking}.  Second, the conversion mechanism into escaping radiation is problematic, referred to as a ``bottle-neck'' by \citet{U00}. Details are discussed in Section~\ref{sect:RPE}.

\noindent
{\bf ADE}: Due to the extremely strong magnetic field in a pulsar plasma, all electrons (and positrons) quickly radiate away the perpendicular component of their energy, so that they are in 1D motion along the field lines. This extreme form of anisotropy is a source of free energy that can drive an anomalous Doppler instability \citep{MU79,LP82,KMM91a,LBM99a}. In this case, the resonance condition for an electron with speed $\beta$ is $\beta/z-1=\Omega_{\rm e}/\omega\gamma$, where $\Omega_{\rm e}=eB/m$ is the electron cyclotron frequency. The major difficulty with ADE is that the frequency is too high: the resonance condition and the wave properties require $\omega\gg\Omega_{\rm e}/\gamma$. We argue that this condition is too restrictive for ADE to be plausible as the generic pulsar radio emission mechanism (details in Section~\ref{sect:ADE}).

Wave dispersion in a pulsar plasma plays an important role in our discussion of possible radio emission mechanisms. In three recent papers, referred to here as RMM1 \citep{RMM19a_2019JPlPh..85c9005R}, RMM2 \citep{RMM19b_2019JPlPh..85c9011R} and RMM3 \citep{RMM19_3} we discussed aspects of the plasma physics relevant to a pulsar plasma in detail. In RMM1 we discussed wave dispersion in the rest frame of a pulsar plasma. In RMM2 we argued that the widely-made choice of a relativistically streaming Gaussian (RSG) distribution is artificial, and that a more realistic choice involves starting with an appropriate distribution in the rest frame, Lorentz-transforming this distribution and identifying the streaming distribution as this Lorentz-transformed distribution (LTD). The LTD is very much broader than any RSG in the highly relativistic limit. This assumption underlies our negative conclusions (RMM3) concerning the efficacy of beam-driven wave growth. A notable difference that we identify for a model based on LTD, compared with a model based on RSG, concerns the conclusion that wave growth is due the reactive
version of the beam-driven instability, because the growth rate of the kinetic instability would exceed the bandwidth of the growing waves \citep{ELM83}. In contrast we (RMM3) found that the much broader form of a LTD model leads to  much smaller growth rate for the kinetic instability and a much larger bandwidth of the growing waves, such that this inequality is reversed; we also found that the large bandwidth precludes the existence of a reactive version of the instability.
Another notable consequence of a LTD model concerns the requirement that the total distribution of particles, that is the sum of the background and beam distributions, have a well-defined minimum that separates the beam and background distributions, in order for there to be a positive slope in the distribution function (above the minimum) to drive the kinetic instability. This separation condition is much more difficult to satisfy for a LTD model than for a RSG model. We discuss this problem further in Section~\ref{sect:dispersion}.

The properties of wave dispersion in the pulsar plasma, and the instabilities that generate the waves, play a direct role in RPE and ADE and an indirect role in the favored version of CCE. We point out that oversimplified and misleading assumptions relating to the wave dispersion have been made, either explicitly or implicitly, in many existing treatments of the instabilities involved. There is a dichotomy in the literature from the 1970s on beam-driven instabilities in a pulsar plasma between those who assume the plasma to be cold or nonrelativistic in its rest frame and those who took dispersion in the relativistic plasma into account \citep[e.g.,][]{ELM83,AM98,L99,MG99,MGKF99}. On the one hand, the assumption that the plasma is nonrelativistic in its rest frame underlies an (implicit or explicit) assumption that a beam-driven instability causes ``Langmuir-like'' waves to grow. Specifically the waves that grow, in the rest frame of the plasma, are assumed to have properties similar to those of Langmuir waves in a nonrelativistic thermal plasma, notably, frequency near the plasma frequency, $\omega\approx\omega_{\rm p}$, or some relativistic counterpart, longitudinal polarization and phase speeds that can be driven by a nonrelativistic beam. This nonrelativistic assumption continues to be made in some treatments of RPE \citep[e.g.,][]{EH16}. On the other hand, when the plasma is assumed to be relativistic, $\av{\gamma}-1\gtrsim1$ in its rest frame, the properties for the wave dispersion are quite different. The dominating effects of $\av{\gamma}-1\gtrsim1$ on wave dispersion in a pulsar plasma have been recognized since the 1970s \citep[e.g.,][]{KT73,LM79,VKM85,AB86,MG99,LBM99,AR00,MelikidzeMG14}. As discussed in detail in RMM1, all waves in such a plasma have phase speeds that are either just below unity (subluminal), with $\gamma_\phi=(1-z^2)^{-1/2}\gg\av{\gamma}$, or above unity (superluminal). This feature of the wave dispersion is the basis for the statement: there are no Langmuir-like waves in a pulsar plasma.

In Section~\ref{sect:parameters} we discuss the parameters of the pulsar plasma assumed here in treating the wave dispersion.  In Section~\ref{sect:beam} we discuss relativistic relative motions and beam speeds, and the implications for resonance between a wave and a beam. In Section~\ref{sect:dispersion} we summarize the properties of wave dispersion in a pulsar plasma. In Sections~\ref{sect:CCE}, \ref{sect:RPE} and~\ref{sect:ADE} we apply the results to critical discussions of CCE, RPE and ADE, respectively. In connection with the discussion of CCE, we identify three types of coherence mechanism: superradiance, reactive instabilities and kinetic (or maser) instabilities, and argue that none of them can account for the postulated coherence in CCE. In Section~\ref{sect:discussion} we summarize our arguments concerning the viability of the suggested radio emission mechanisms. Our conclusions are summarized in Section~\ref{sect:conclusions}

\section{Parameters for a pulsar plasma}
\label{sect:parameters}

In this section we summarize the assumptions made about the pulsar plasma, and estimate the value of  parameters relevant to the wave dispersion.

\subsection{Reference frames}

In our calculations, three reference frames are of note: the rest frame of the background $ \mathcal{K} $, the pulsar frame $ \mathcal{K}' $ and the rest frame of the beam $ \mathcal{K}'' $. Frame $ \mathcal{K} $ propagates outwards with speed $ \beta_{\rm s} $ and corresponding Lorentz factor $ \gamma_{\rm s} $ with respect to $ \mathcal{K}' $; frame $ \mathcal{K}'' $  propagates outwards with $ \beta_{\rm r}, \gamma_{\rm r} $ with respect to $ \mathcal{K}' $; and $ \mathcal{K}'' $ propagates outwards with $ \beta_{\rm b}, \gamma_{\rm b} $ with respect to $ \mathcal{K} $. One has
\begin{equation}
    \gamma' = \gamma_{\rm s}\gamma(1 + \beta_{\rm s}\beta) = \gamma_{\rm r}\gamma''(1 + \beta_{\rm r}\beta''),\quad
    \gamma'' = \gamma_{\rm b}\gamma(1 - \beta_{\rm b}\beta),
\end{equation}
where a single (double) prime denotes parameters in $ \mathcal{K}' $ ($ \mathcal{K}'' $). In particular, $ \gamma_{\rm s} = \gamma_{\rm r}\gamma_{\rm b}(1 + \beta_{\rm r} \beta_{\rm b}) \approx 2\gamma_{\rm r} \gamma_{\rm b} $, where the approximation applies for $ \gamma_{\rm r}, \gamma_{\rm b} \gg 1 $. In RMM3 we show that for maximum growth through weak-beam instability we require $ \gamma_{\rm b} \approx (10{\rm -}20)\av{\gamma} $, where $ \av{\gamma} $ is the average spread of the background distribution in its rest frame, with the averages defined as in RMM2.

\subsection{Pulsar plasma}

In a polar-cap model \citep[e.g.,][]{Michel91,BGI93,M99,LG-S06} the  source of the radio emission is assumed to be on open field lines in a relativistically outflowing electron/positron plasma created by pair cascades, referred to here as a ``pulsar plasma''. In early models charges were assumed to be drawn from the stellar surface in the polar-cap regions, and accelerated to very high energy by a parallel electric field, $E'_\parallel$
in a vacuum gap or double layer above the surface \citep{GJ69,RS75,A79,BGI86}. Such particles were referred to as ``primary'' particles and they were assumed to be accelerated  by $E'_\parallel$ to Lorentz factors $\gamma' =10^6$--$10^7$ where the acceleration is balanced by energy loss through curvature radiation \citep{UM95}. The curvature emission produces $\gamma$-rays, which decay into ``secondary'' electron-positron pairs in the superstrong magnetic field. The secondary particles are further accelerated by $E'_\parallel$, producing more $\gamma$-rays until the resulting pair cascade \citep{HA01,AE02,ML10} results in a dense enough plasma to screen $E'_\parallel$ above the gap or double layer, to maintain the charge density at close to the corotation value $\rho'_{\rm cor}$.
{\br An important ingredient in the discussion here is that pair recreation is intermittent and intrinsically time-dependent, as pointed out by \citet{Beskin_82}, cf. also \citep{Letal05,B08,Lyubarsky09,T10b}, rather than pair recreation proceeding in a steady state as assumed, for example, in the carousel model for subpulses \citep{RS75}. 
}
Particle-in-Cell calculations \citep[e.g.,][]{TA13}, have not led to radically different models for the (time-averaged) properties of the bulk of the  pair plasma, compared with these earlier models. 

{\br The assumption that primary particles originate from the stellar surface was questioned by \citet{Beskin_82}. One suggested alternative is that the primaries may be cosmic rays that penetrate into the inner magnetosphere \citep{SR82}. We note that models that take the intrinsic time-dependence into account and result in large-amplitude oscillations \citep[e.g.,][]{Letal05,B08,Lyubarsky09,T10b} do not rely on primaries from the stellar surface. 
}

Here we assume that ``plausible'' parameters for the pulsar plasma correspond to pairs streaming outward with a bulk Lorentz factor $\gamma_s$ of order $10^2$--$10^3$, and with an intrinsic relativistic spread with $\langle\gamma\rangle$ between a few and about 10  \citep{HA01,AE02}. Numerical models for pair cascades also imply the ratio between the number density of pairs and $\rho_{\rm cor}/e$, referred to as the multiplicity factor, $\kappa$. {\br The value of $\kappa$ was estimated to be $ 10^5$ by \citet{TH15}, and \citet{TH19} estimated the maximum to be about $10^6$; \citet{BGI93} estimated a smaller value, $\kappa=10^3$--$10^4$. We adopt the fiducial value $\kappa=10^5$.} We further quantify what we mean by ``plausible'' parameters for the pulsar plasma in the remainder of this section.

\subsection{Plasma parameters}
\label{section:plasma_parameters}

We characterize the plasma by three plasma parameters: the electron cyclotron frequency, $\Omega_{\rm e}=eB/m$, the plasma frequency, $\omega_{\rm p}=(e^2n/\varepsilon_0m)^{1/2}$ and the ratio, $\beta_{\rm A}$, of the Alfv\'en speed to the speed of light. No Lorentz factors are included in our definitions of $\Omega_{\rm e}$, $\omega_{\rm p}$ and $\beta_{\rm A}$.
The number density $n'$ in ${\cal K}'$ is related to $n$ in ${\cal K}$ by $n'=\gamma_{\rm s}n$. As conventionally defined the Alfv\'en speed is $v_{\rm A}=\beta_{\rm A}c=B/(\mu_0nm)^{1/2}$ in ${\cal K}$, and this is much greater than the speed of light, $\beta_{\rm A}\gg1$, in a pulsar plasma.  We need estimates of $\Omega_{\rm e}$, $\omega_{\rm p}$ and $\beta_{\rm A}$ as functions of the radial distance, $r$, with $r$ referred to as the ``height'' where no confusion should result. 

An estimate of $\Omega_{\rm e}$ follows from the (polar) magnetic field at the surface of the star, $B_* = 3.2\times 10^{15}(P{\dot P})^{1/2}\,$T, where $ P $ is the pulsar period and $ \dot{P} $ is the period derivative, together with the dipole approximation implying $B=B_*(R_*/r)^3$ for $R_*<r\ll r_{\rm LC}$, where $R_*\approx10^4\,$m is the radius of the star, and $r_{\rm LC}=Pc/2\pi$ is the light cylinder radius. It is convenient to write the dependence on $r$ in terms of either the ratio $r/r_{\rm LC}\ll1$ or the ratio $r/R_*\gg1$. The plasma frequency in the pulsar (primed) frame can be estimated assuming that the electron density is greater than the corotation charge density (divided by the fundamental charge $e$)  by the multiplicity factor $\kappa$. This gives $\omega'^2_{\rm p}\approx\kappa\Omega_*\Omega_{\rm e}$ in ${\cal K}'$, with $\Omega_*=2\pi/P$ the rotation frequency of the star, implying $\omega_{\rm p}^2\approx\kappa\Omega_*\Omega_{\rm e}/\gamma_{\rm s}$ in ${\cal K}$. 

As fiducial values we assume $P=1\,$s and ${\dot P}=10^{-15}$ for a normal pulsar, giving ${\dot P}/P^3=10^{-15}\,{\rm s}^{-3}$. The value of ${\dot P}/P^3$ is relatively insensitive to the variation in $P$ and ${\dot P}$ between recycled pulsars, normal pulsars and magnetars. We further adopt the fiducial values $\kappa=10^5$, $\av{\gamma}=10$ and $\gamma_{\rm s}=10^3$. These values give, in ${\cal K}$,
\begin{equation}\label{parameters}
\begin{split}
    \frac{\Omega_{\rm e}}{2\pi} 
        & \approx 26{\rm\,GHz}\left(\frac{{\dot P}/P^3}{10^{-15}\,{\rm s}^{-3}}\right)^{1/2}\left(\frac{r/r_{\rm LC}}{0.1}\right)^{-3}\left(\frac{1\,{\rm s}}{P}\right),\\
    \frac{\omega_{\rm p}}{2\pi} 
        & \approx 1.6{\rm\,MHz}
        \left[\left(\frac{\kappa}{10^5}\right)
        \left(\frac{10^3}{\gamma_{\rm s}}\right)
        \left(\frac{{\dot P}/P^3}{10^{-15}\,{\rm s}^{-3}}\right)^{1/2}\left(\frac{r/r_{\rm LC}}{0.1}\right)^{-3}\right]^{1/2}\left(\frac{1\,{\rm s}}{P}\right),\\
    \beta_{\rm A}^2 
        & \approx 2.6\times10^7\left(\frac{10}{\av{\gamma}}\right)\left(\frac{10^5}{\kappa}\right)
        \left(\frac{\gamma_{\rm s}}{10^3}\right)\left(\frac{{\dot P}/P^3}{10^{-15}\,{\rm s}^{-3}}\right)^{1/2}\left(\frac{r/r_{\rm LC}}{0.1}\right)^{-3}.
\end{split}
\end{equation}
The height $r=0.1r_{\rm LC}$ is close to the maximum usually considered possible; a height of several tens of stellar radii is considered more plausible, e.g., for $r = 30R_* $ one has $ r/r_{\rm LC} \approx 6.3\times10^{-3}/P $.

\begin{figure}
\centering
\psfragfig[width=1.0\columnwidth]{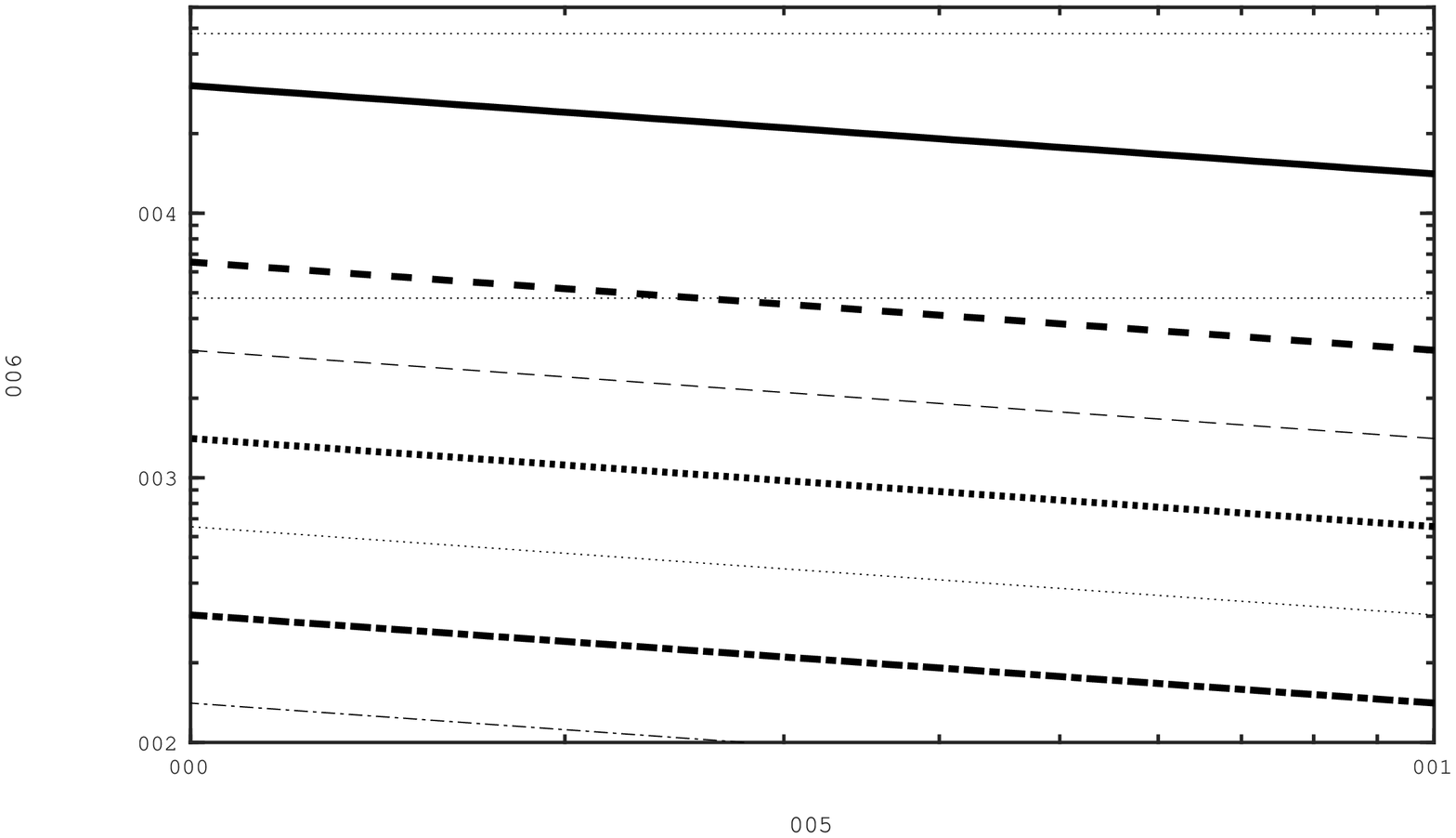}\\\vspace{2mm}
\psfragfig[width=1.0\columnwidth]{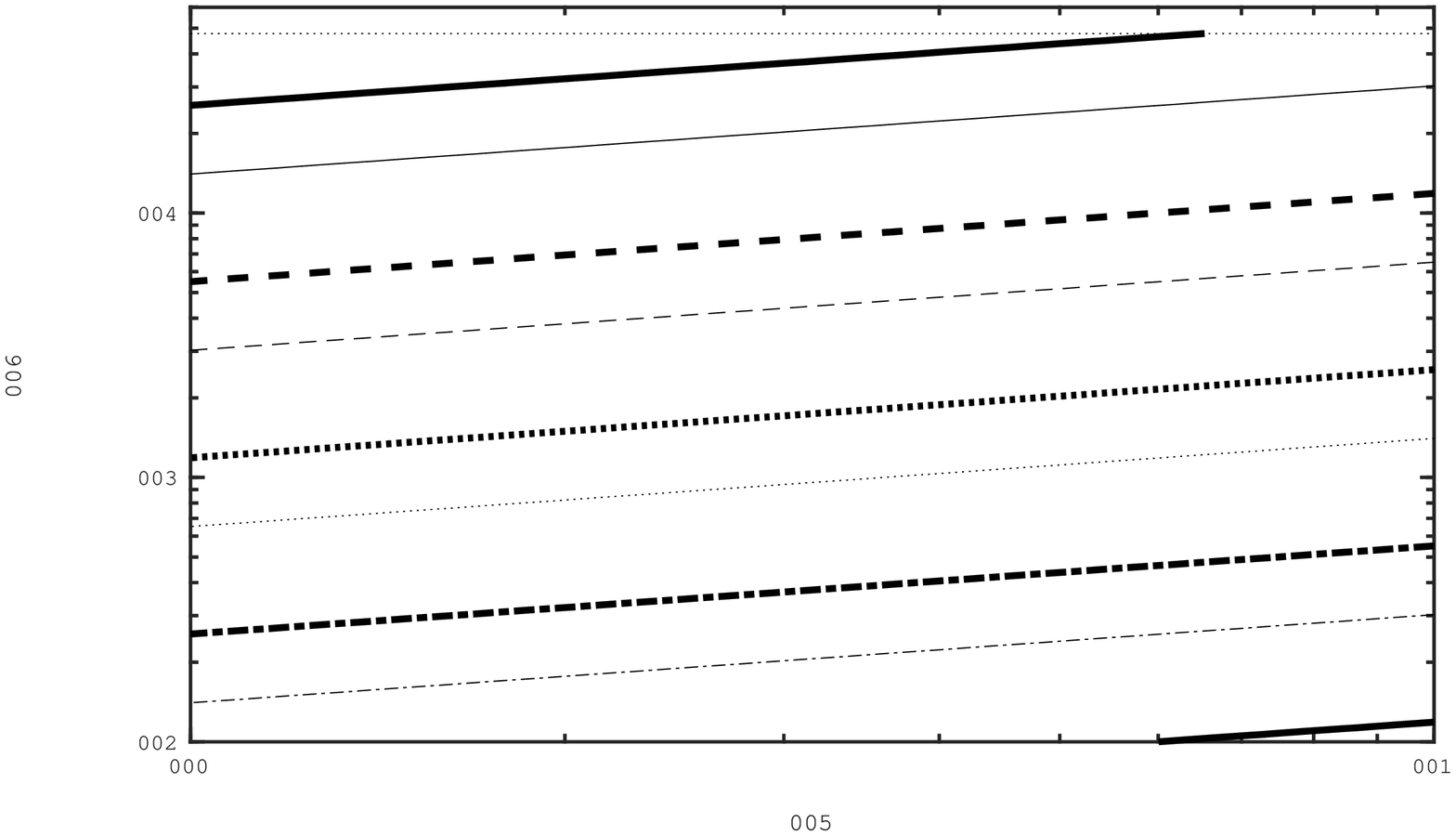}
\caption{TOP: Contour plots of $ \omega_{\rm p}/2\pi $ as a function of $ \gamma_{\rm s} $ and $ r/R_* $ for $ P = 1\, {\rm s} $ (thick) and $ P = 0.1\, {\rm s} $ (thin). We use $ \kappa = 10^5 $ and $ \dot{P}/P^3 = 10^{-15}\,{\rm s}^{-3} $ for all plots. The contours are at $ \omega_{\rm p}/2\pi = 10^7\rm\,$Hz (solid), $ 10^8\rm\,$Hz (dashed), $ 10^9\rm\,$Hz (dotted) and $ 10^{10}\rm\,$Hz (dash-dotted). The thin dotted horizontal lines are at $ r/r_{\rm LC} = 0.1 $ for $ P = 1\,{\rm s} $ (upper) and $ P = 0.1\,{\rm s} $ (lower); and $ r/R_* = 1 $ indicates the stellar surface. BOTTOM: Contour plots of $ \beta_{\rm A} $ as a function of $ \gamma_{\rm s} $ and $ r/R_* $ for $ \av{\gamma} \approx 1.7 $ (thick) and $ \av{\gamma} \approx 10 $ (thin). We use $ \kappa = 10^5 $ and $ \dot{P}/P^3 = 10^{-15}\,{\rm s}^{-3} $ and $ P = 1\,{\rm s} $ for all plots. The contours are at $ \beta_{\rm A} = 10^4 $ (top solid), $ 10^5 $ (dashed), $ 10^6 $ (dotted), $ 10^7 $ (dash-dotted) and $ 10^8 $ (bottom solid). The thin dotted horizontal line is at $ r/r_{\rm LC} = 0.1 $ and $ r/R_* = 1 $ indicates the stellar surface.}
\label{fig:omega_p}  
\end{figure}

Figure~\ref{fig:omega_p} shows contour plots of $ \omega_{\rm p}/2\pi $ (TOP) and $ \beta_{\rm A} $ (BOTTOM). We use $ \kappa = 10^5 $ and $ \dot{P}/P^3 = 10^{-15}\,{\rm s}^{-3} $ for all plots. TOP: Contour plots of $ \omega_{\rm p}/2\pi $ as a function of $ \gamma_{\rm s} $ and $ r/R_* $ for $ P = 1\, {\rm s} $ (thick lines) and $ P = 0.1\, {\rm s} $ (thin lines). The contours are at $ \omega_{\rm p}/2\pi = 10\rm\,$MHz (solid), $ 100\rm\,$MHz (dashed), $ 1\rm\,$GHz (dotted) and $ 10\rm\,$GHz (dash-dotted). The thin dotted horizontal lines are at $ r/r_{\rm LC} = 0.1 $ for $ P = 1\,{\rm s} $ (upper) and $ P = 0.1\,{\rm s} $ (lower); and $ r/R_* = 1 $ indicates the stellar surface. BOTTOM: Contour plots of $ \beta_{\rm A} $ as a function of $ \gamma_{\rm s} $ and $ r/R_* $ for $ \av{\gamma} \approx 1.7 $ (thick lines) and $ \av{\gamma} \approx 10 $ (thin lines). The contours are at $ \beta_{\rm A} = 10^4 $ (top solid), $ 10^5 $ (dashed), $ 10^6 $ (dotted), $ 10^7 $ (dash-dotted) and $ 10^8 $ (bottom solid). The thin dotted horizontal line is at $ r/r_{\rm LC} = 0.1 $ and $ r/R_* = 1 $ indicates the stellar surface.

Another parameter that appears is the radius of curvature, $R_c$, of the magnetic field lines. For the polar-cap model, an approximate estimate \citep{KL10} is
\begin{equation}
    R_c 
        \approx (rr_{\rm LC})^{1/2} 
        \approx 1.5\times10^7 \,{\rm m}\left(\frac{P}{1\,{\rm s}}\right)\left(\frac{r/r_{\rm LC}}{0.1}\right)^{1/2}.
\end{equation}

{\br The plausible range of these parameters is determined by possible ranges about our chosen fiducial values. As already noted, the possible range of $\kappa$ is between $10^3$ and $10^6$, with the values of $\omega_p$ and $\beta_{\rm A}$ depending on the square root of this parameter. The plausible range of $\langle\gamma\rangle$ is $\approx2$ to $\approx10$; only $\beta_{\rm A}$ depends on (the square root of) this parameter. The height is the most uncertain of the assumed parameters, with $r=0.1r_L$ close to the maximum usually considered possible. A height of several tens of stellar radii is considered more plausible; one may reflect this by making the alternative choice of $r/R_*=30$ as the fiducial value, with the values of the parameters for this choice modified by those given by the factor $(r/0.1r_L)^{-3/2}\approx63$.
}

\subsection{Source height}

The source region of the radio emission is uncertain, particularly the emission height \citep{GG03,Detal04,KJ07}. For example, \citet{Jetal08} argued that the emission height changes from high in young pulsars to low in older pulsars, with emission from a broad range of heights for intermediate ages. More recently, \citet{Mitra17} summarized three different ways of determining the height from observational data, and concluded that the source height is at $r/r_{\rm LC}<0.1$. With $r/r_{\rm LC}\le0.1$, our fiducial values give $\Omega_{\rm e}/2\pi\ge26\rm\,GHz$,  $\omega_{\rm p}/2\pi\ge1.6\rm\,MHz$, and $\beta_{\rm A}\ge5.1\times10^3$. 

The various specific estimates of the height mostly give values of $r/R_*$ between several tens and a few hundreds, which corresponds to $r/r_{\rm LC}$ between several $10^{-3}/P$ and several $10^{-2}/P$. In particular, for $P=1\,$s and $R_*=10^4\,$m, $r/R_*=30$ correspond to $(r/0.1r_{\rm LC})^{-3/2}\approx63$. We note that for $r=30R_*$ and the values of $\kappa$, $P$, ${\dot P}$ as in~(\ref{parameters}), $\omega_{\rm p}/2\pi$ is approximately $100\,$MHz, and $\beta_{\rm A}$ is approximately $3.2\times10^5$. 

\section{Beam-driven resonant waves}
\label{sect:beam}

In this section we comment on suggested models for the formation of beams in a pulsar plasma, and then discuss some implications of wave-particle resonance involving a beam.

\subsection{Possible beams}
\label{sect:possible}

In the early literature on RPE two different types of beams were considered: a beam of primary particles moving through secondary pair plasma, and  relative motion of electrons and positrons associated with the pulsar current. The primary particles were assumed to have very high bulk outflow Lorentz factors, $\gamma_{\rm p}=10^6$--$10^7$, and number density $n'_{\rm p}$, with comparable energy density is the primary and secondary particles, $\gamma_{\rm p}n'_{\rm p}\approx\gamma_{\rm s}n'$, where $ \gamma_{\rm s} $ is the bulk outflow Lorentz factor of the secondary particles. The relative motion of electrons and positrons is required for the current density needed to satisfy the electrodynamics. Neither model can account for the required wave growth  \citep[e.g.,][]{Letal86}. This led to the suggestion of a multiple-sparking model, in which the production of the secondary pair plasma, through pair cascades, results in localized transient ``clouds'' of pair plasma \citep{U87,U02,UU88,AM98}. The name ``multiple-sparking'' applies to an older version of the model in which the source of the primary particles was assumed to be favored locations (sometimes called ``hot spots'') on the stellar surface \citep{RS75,FR_82,Beskin_82,Gil_S00}. In more recent models, in which the intrinsic time-dependence is taken into account \citep[e.g.,][]{T10b,TA13}, the pair creation exhibits a limit cycle behavior that could be considered similar to what is assumed in a sparking model. We use the name ``multiple-beam'' to refer to any model in which pair cascades result in  localized transient clouds. 

\subsection{Multiple-beam model}
\label{sect:multiple-sparking}

A widely favored model for the formation of multiple beams involves faster particles in a ``trailing'' cloud overtaking slower particles in a ``leading'' cloud  \citep{U87,U02,UU88,AM98}. Once the overtaking has occurred, the faster particles from the trailing cloud may be regarded as a beam propagating through the slower particles in the leading cloud.
Here we discuss the efficacy of this model critically, first by considering a model proposed by \cite{AM98}, and then based on a model developed in the Appendix. 

The multi-beam model proposed by \cite{AM98} involves multiple clouds of pairs, with each cloud postulated to be initially of length $L'_0$ with a gap initially of length $h'_0$ separating sequential clouds; they chose $h'_0=100\,$m and $L'_0=(30{\rm-}40)h'_0 = (3{\rm-}4)\times10^3$\,m. \cite{AM98} separated the electrons in each cloud into three speeds, called fast, intermediate and slow. We simplify the model by considering only fast and slow electrons, denoted F and S, respectively. Beam formation is attributed to F~particles in a trailing cloud overtaking S~particles is the preceding leading cloud. This occurs after a time 
\be
t'_{\rm FS} = \frac{L'_0 + h'_0}{c(\beta'_{\rm F} - \beta'_{\rm S})} 
            \approx \frac{L'_0+h'_0}{c}\frac{2\gamma'^2_{\rm F}\gamma'^2_{\rm S}}{\gamma'^2_{\rm F}-\gamma'^2_{\rm S}},
\label{tFS}
\ee
where $ L'_0 + h'_0 $ is the initial separation of particles. For illustration purposes, \cite{AM98} chose $\gamma'_{\rm F}=300$, $\gamma'_{\rm S}=100$. These numbers give $t'_{\rm FS}\approx0.2\,$s for the time required for overtaking to occur. 

A serious difficulty with this model is that $0.2\,$s is too long. Specifically, in $0.2\,$s no beam could form (a) inside the light cylinder for a pulsar with period $ P < 1.3 $\,s
or (b) inside $r/r_{\rm LC}=0.1$ for $ P < 12.6 $\,s.
We conclude that with these numbers, the overtaking-cloud model cannot lead to effective beam formation in most pulsars.

This difficulty is further compounded when one takes a plausible values $\av{\gamma}\lesssim10$ for the intrinsic spread in Lorentz factors resulting from pair cascades. For a uniform distribution, the choice $\gamma'_{\rm F}=300$, $\gamma_{\rm S}=100$ would require $\av{\gamma}\approx100$, with $\gamma'_{\rm F}\approx200+\av{\gamma}$, $\gamma_{\rm S}\approx200-\av{\gamma}$. With $\av{\gamma}\approx10$ a more appropriate choice would be $\gamma'_{\rm F}\approx210$, $\gamma'_{\rm S}\approx190$. With these revised numbers one has $t'_{\rm FS}\approx 4\,$s. This is an impossibly long time for beam formation to be relevant. This has a simple explanation: the smaller the difference in Lorentz factors between the fast and slow particles, the smaller is their relative speed, and hence the longer it takes for a fast particle to catch a slow particle. For a J\"uttner distribution, truncated as discussed in the Appendix, the requirement becomes that $ \av{\gamma^2} \approx 10/9 $ which corresponds to an extremely cool plasma whereas in pulsars we have $ 200 \gtrsim \av{\gamma^2} \gtrsim 4 $ for $ 9 \gtrsim \av{\gamma} - 1 \gtrsim 1 $. For the nominal value of $ \av{\gamma} = 10 $, for a truncated J\"uttner distribution we have $ \gamma'_{\rm S} \approx \gamma_{\rm s}/80 $ and $ \gamma'_{\rm F} \approx 80\gamma_{\rm s} $ with $ \gamma_{\rm s} = 10^2{\rm -}10^3 $. 

\subsection{Fractionization}

The model of \cite{AM98} does not allow one to discuss fractionation in any detail. Consider a given cloud that initially has a uniform distribution of particles within a cylinder of length $L'_0$. As the beam propagates, its length increases and it becomes increasingly inhomogeneous, in the sense that the distribution function at any point in the beam becomes narrower, with the local (at a given location along the beam) average speed decreasing from the front to the back of the beam. It is this effect that we refer to as fractionation. 

A simple ballistic model suffices to describe how fractionization occurs. Suppose that all particles are confined to $-L'_0/2<x<L'_0/2$ at $t=0$. After a time $t$ when particles with velocity $v_\parallel$ have traveled a distance $d=v_\parallel t$, particles with velocities $v_\parallel\pm\Delta v_\parallel/2$ have traveled an additional distance $\pm \Delta v_\parallel t/2$. Particles with $v_\parallel\pm\Delta v_\parallel/2$ become spatially separated from each other when the difference between these two additional distances exceeds $L'_0$. It follows that after propagating a distance $d\gg L'_0$ the particles at a given point within the extended beam are confined to a range
\begin{equation}
\Delta v_\parallel=\frac{L'_0}{d}v_\parallel,
\end{equation}
where $v_\parallel$ may be approximated by the beam velocity $v_{\rm b}$.

A more detailed discussion of fractionization and its implications is given in Appendix~\ref{app:convection}, where we raise the possibility that once overlapping starts the local distribution function may have two narrow peaks, a slower one from the original leading beam, and a faster one from the original trailing beam. Such a two-peaked distribution may lead, in principle, to reactive growth of waves, but with significant changes to the usual model, including the need for the beam to be the slower leading cloud. 

We conclude that the conditions for overtaking in a multiple-beam model are considerably more complicated than has been recognized in existing discussions of the model. Increasing length and fractionation need to be taken into account in both the leading and trailing beams, and what ``overtaking'' means needs to be defined. For most parameters considered plausible a realistic form of overtaking does not occur during the time it takes for the beams to propagate from the stellar surface to a plausible source height for the radio emission. Although we doubt that the overtaking-cloud model is viable at all, we ignore this difficulty in the following discussion, postulating the overtaking might occur and consider the implications for beam-driven wave growth. 

\subsection{Relativistically streaming distributions}

In pulsar plasma the Lorentz factors that describe the intrinsic spread, $\av{\gamma}$ in the rest frame, and the outward streaming, $\gamma_{\rm s}$, are assumed to satisfy $\gamma_{\rm s}\gg \av{\gamma} - 1 \gtrsim 1 $. In RMM2 we showed that in any such model, the spread in Lorentz factors in the pulsar frame, in which this plasma is streaming, is very much larger than $\av{\gamma}$. Before discussing more general distributions, we show this to be the case for a ``water-bag'' model for the distribution distribution function $g(u)$ as a function of 4-speed $u=\gamma\beta$: 
\be
    g(u) = 
    \begin{dcases}
        n/2u_1, & \quad \abs{u} < u_1,\\
        0, & \quad \text{otherwise}.
    \end{dcases}
\label{waterbag}
\ee
where $n$ is the number density in the rest frame. Assuming $u_1\approx\gamma_1\gg1$, the mean Lorentz factor is $\av{\gamma}\approx\gamma_1$ in the rest frame. The spread in Lorentz factors is $1\le\gamma\le\gamma_1$. In the pulsar (primed) frame, $\pm u_1$ transform to $u'_\pm=\gamma_{\rm s}\gamma_1(\beta_{\rm s} \pm \beta_1)$, or $u'_+\approx2\gamma_{\rm s}\gamma_1$, $u'_-\approx\gamma_{\rm s}/2\gamma_1$. The spread in Lorentz factors in this frame is approximately $\gamma_{\rm s}\gamma_1$ which is much greater (by a factor of order $\gamma_{\rm s}$) than the spread $\av{\gamma}\approx\gamma_1$ in the rest frame.

A widely favored choice for the distribution function of a beam is a relativistically streaming Gaussian (RSG) of the form
\be 
g_{\textsc{RSG}}(u)\propto\exp\left[-{(u-u_{\rm s})^2}/{u_{\rm T}^2}\right],
\label{RSG}
\ee
where $u_{\rm s}=\gamma_{\rm s}\beta_{\rm s}$ is the streaming 4-speed, and $u_{\rm T}$ may be interpreted as the spread in 4-speed about $u=u_{\rm s}$. The RSG does not retain its form under a Lorentz transformation. For example, the Lorentz transformation to the rest frame of the distribution, denoted by a tilde, implies that $\beta$ transforms to ${\tilde\beta}=(\beta-\beta_{\rm s})/(1-\beta\beta_{\rm s})$ and the distribution function transforms to ${\tilde g}_{\textsc{RSG}}({\tilde u})=g_{\textsc{RSG}}(u)$ with $u-u_{\rm s}=({\tilde\gamma}-1)u_{\rm s}+\gamma_{\rm s}{\tilde u}$. This rest-frame distribution has its maximum at ${\tilde u}=0$ or ${\tilde\beta}=0$, but it is not a symmetric function of ${\tilde u}$ or ${\tilde\beta}$, and $u_{\rm T}$ cannot be interpreted as the spread in ${\tilde u}$. 

We suggest that the choice of a RSG distribution is artificial, and is made primarily for mathematical convenience.

\subsection{Including streaming by a Lorentz transformation}

We argue that the appropriate choice for a relativistically streaming distribution is that obtained by a applying a Lorentz distribution to a plausible rest-frame distribution, e.g., to a J{\"u}ttner distribution or a Gaussian distribution. This procedure results in Lorentz-transformed J{\"u}ttner (LTJ) and a Lorentz-transformed Gaussian (LTG) distribution,
\be 
g'_{\textsc{LTJ}}(u')\propto\exp\left[-\rho\gamma'\right],
\qquad
g'_{\textsc{LTG}}(u')\propto\exp\left[-{u'^2}/{u_{\rm T}^2}\right],
\label{RSGp}
\ee
respectively, with $\gamma'=\gamma\gamma_{\rm s}(1-\beta\beta_{\rm s})$ and $u'=\gamma\gamma_{\rm s}(\beta-\beta_{\rm s})$. (A third example is a Lorentz-transformed water-bag distribution, cf. the discussion following equation (\ref{waterbag}).) We are concerned with the highly relativistic case in which both the streaming is highly relativistic, $\gamma_{\rm s}\gg1$ and the spread in the rest frame is (highly) relativistic, $\av{\gamma} - 1 \gtrsim 1 $, with $\av{\gamma} \approx 1/\rho $ for a J{\"u}ttner distribution, and $\av{\gamma} \approx u_{\rm T}$ for a Gaussian distribution. The negative exponents in the RSG, LTJ and LTG may then be approximated by
\begin{equation}
    \frac{(\gamma-\gamma_{\rm s})^2}{\av{\gamma}^2},
    \qquad 
    \frac{(\gamma-\gamma_{\rm s})^2}{2\gamma\gamma_{\rm s}\av{\gamma}},
    \qquad
    \left(\frac{(\gamma-\gamma_{\rm s})(\gamma+\gamma_{\rm s})}{2\av{\gamma}\gamma\gamma_{\rm s}}\right)^2,
\end{equation}
respectively.
It follows that the LTJ and LTG distributions are broader than a RSG distribution by of order $2\gamma_{\rm s}$ and $\gamma_{\rm s}^2$, respectively. This surprising (at least to us) result implies that the choice of a RSG is misleading in that it can lead to a serious underestimate of the spread in Lorentz factors for a relativistically streaming distribution obtained by Lorentz transforming a rest-frame distribution. 

Replacing the RSG by LTJ or LTG leads to a large increase in the spread in Lorentz factors, from $\av{\gamma}$ in the rest frame  of the distribution, to of order $\gamma_{\rm s}\av{\gamma}$ for the Lorentz-transformed distribution in the pulsar frame. This is a characteristic feature of any model in which relativistic streaming is included by Lorentz transforming. With our ``plausible'' parameters, a spread of order $\av{\gamma}\approx10$ is increased to $\gamma_{\rm s}\av{\gamma} \sim 10^3{\rm-}10^4$, where we use $ \gamma_{\rm s} \sim 10^2{\rm-}10^3$. We conclude that this is a potentially very large effect that cannot be ignored. The choice of a RSG does ignore this effect. As discussed below, estimates of the growth rate and of the bandwidth of the growing waves depend strongly on the width of the relativistic streaming distribution. Moreover, we found that the condition for a beam-driven reactive instability to exist is not satisfied (RMM3).

\subsection{Separation condition}

The inclusion of streaming by applying a Lorentz transformation to a non-streaming distribution function has a large effect on the separation condition for the two distributions.
A requirement for beam-driven instability to develop is that the beams (or the beam and the background in a weak-beam model) do not overlap significantly in momentum space, e.g., in $\gamma$. This requirement is the separation condition.

In RMM2 we showed that for two counter-streaming distributions with equal densities and equal spreads, $\av{\gamma}_1=\av{\gamma}_2\to\av{\gamma}$, the two beams become separated in the frame in which they are counter-streaming when the Lorentz factor of the counter streaming exceeds about $\av{\gamma}$, as one might anticipate.  When this separation condition is Lorentz transformed to the rest frame of one distribution, with the other streaming relative to it at $\gamma_{\rm b}$, this condition transforms into $\gamma_{\rm b}>2\av{\gamma}^2$.\footnote{In a weak-beam model the separation condition is $\gamma_{\rm b} \gtrsim 10\av{\gamma}$ which require larger $ \gamma_{\rm b} $ for distributions with $ \av{\gamma} < 10 $ than that for equal beams counter streaming.}
This separation condition applies to any distribution with $\av{\gamma} - 1 \gtrsim 1$ when it is Lorentz-transformed to become a streaming distribution, cf. (\ref{RSGp}). 

In contrast, when a RSG distribution is chosen, the separation condition is much more easily satisfied because the spread in each distribution is much smaller, e.g., by a factor of order $1/\gamma_{\rm s}$.  The choice of a RSG distribution applies only in a single frame. As shown above, a narrow spread in one frame is not preserved under a Lorentz transformation. It is implausible to assume narrow spreads in two independent frames moving relativistically relative to each other.

\subsection{Resonance conditions}

There are two relevant resonance conditions: the Cerenkov condition for a beam instability and the anomalous Doppler condition for ADE. Either can be satisfied only for subluminal waves. 

In general, the gyroresonance condition is
{\br 
\be
\omega-s\Omega_e/\gamma-k_\parallel v_\parallel=0,
\label{gyrores}
\ee
}
with $s=0,\pm1,\ldots$.  In the notation used here the gyroresonance condition becomes
\be
\frac{z-\beta}{z}=s\frac{\Omega_{\rm e}}{\gamma\omega}.
\label{gyroresonance}
\ee
We are interested in resonances at $s\le0$, which require $z\le\beta<1$, where we assume $\beta>0$.\footnote{The resonance condition is written in the rest frame of the plasma. For the distributions discussed here we symmetry about $ z, \beta = 0 $. We could write the resonance condition for $ \beta < 0 $ which would require waves with $ z < 0 $.}

The Cerenkov resonance, $s=0$, requires $z=\beta$ or $\gamma_\phi=\gamma$ in ${\cal K}$ and $z'=\beta'$ or $\gamma'_\phi\approx2\gamma_{\rm s}\gamma$ in ${\cal K}'$, where $ \gamma_\phi = (1 - z^2)^{-1/2} $ is the Lorentz factor corresponding to (subluminal) phase velocity $ z $. The anomalous Doppler resonance, $s=-1$, requires $\beta-z=z\Omega_{\rm e}/\omega\gamma$ in ${\cal K}$ or $\beta'-z'=z'\Omega_{\rm e}/\omega'\gamma'$ in ${\cal K}'$, requiring $|z|<1$ and $|z'|<1$, respectively.

\section{Wave dispersion in pulsar plasma}
\label{sect:dispersion}

In this section we summarize the properties of wave dispersion in a pulsar plasma, both in the rest (unprimed) frame ${\cal K}$ (RMM1) and in the pulsar (primed) frame ${\cal K}'$ (RMM2).

\subsection{RPDF}

Wave dispersion in a pulsar plasma \citep[e.g.,][]{MG99,MGKF99} has two important differences from wave dispersion is a non-relativistic magnetized plasma. First, the Alfv\'en speed is extremely large, $\beta_{\rm A}\gg1$, cf.~\eqref{parameters}; this parameter appears in the wave properties in the combination
{\br
\be
z_A=\beta_A/(1+\beta_A^2)^{1/2}\approx1-1/2\beta_A^2.
\label{zA}
\ee
}
Second, the parallel response involves a relativistic plasma dispersion function (RPDF), which we write as $z^2W(z)$. For a distribution with $\av{\gamma}\gg1$, the real part, $z^2\Re W(z)$, of the RPDF is very sharply peaked, with positive peaks at $z=\pm z_m$ with $z_m^2W(z_m)=2.7\av{\gamma}$, corresponding to $\gamma_\phi=(1-z^2)^{-1/2}$ equal to $\gamma_m=(1-z_m^2)^{-1/2}\approx6\av{\gamma}$. Between these peaks $z^2\Re W(z)$ becomes negative, for $-z_0<z<z_0$, corresponding to $\gamma_\phi<\gamma_0 \approx 1.9\av{\gamma}$, and beyond the peaks, $z^2\Re W(z)$ decreases monotonically with increasing $z>z_m$, being $\approx2\av{\gamma}$ at the light line, $z=1$, and approaching $\langle1/\gamma^3\rangle \approx 1/\av{\gamma}$ for $z\to\infty$. The imaginary part, $z^2\Im W(z)$, of the RPDF is strictly zero in the superluminal range, $z>1$, and it is large, implying strong Landau damping, in the range $z_0\lesssim z\lesssim z_m$.

The foregoing results are derived specifically for a J{\"u}ttner distribution with $\rho\ll1$, for which there is a characteristic scaling, of $z_0,z_m$ etc., with $\rho\approx1/\av{\gamma}$. For $\rho\approx1$, which corresponds to $ \av{\gamma} \approx 1.7 $, this scaling applies with only minor changes in the specific numbers (RMM1). It is only for $\rho\gg1$ that the exact form of $z^2W(z)$ approximates the familiar plasma dispersion function for a thermal plasma; and with $ z^2\Re W(z) \to 1 $ as $ \rho \to \infty $ corresponding to a cold plasma distribution.

Waves of relevance for a resonant instability at $s=0$ (Cerenkov) or $s=-1$ (Doppler) must be subluminal, and subluminal waves exist only in the range $z_0<z<1$. Moreover, the waves in
the range $z_0<z\lesssim z_m$ are strongly (Landau) damped and are ignored here.\footnote{These waves have anomalous dispersion, implying unusual properties including negative energy and superluminal group speed.} Hence the only relevant waves are in the range $z_m\lesssim z<1$ or $\gamma_\phi \gtrsim6\av{\gamma}$. The parameter $\beta_{\rm A}$ is also assumed to be in this range, $\beta_{\rm A} \gg 6\av{\gamma}$.

\begin{figure}
\centering
\psfragfig[width=1.0\columnwidth]{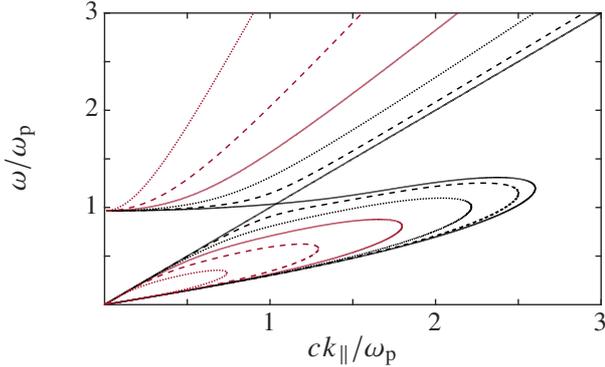}
\caption{Dispersion curves for a nonrelativistic 1D J{\"u}ttner distribution with a temperature $3\times10^8\,$K. The solid black curves correspond to the L~and A~modes for $\theta=0$, and the other nested curves are for the O~mode (upper left) and the Alfv\'en mode (lower right) with $\theta$ increasing in steps of $0.25\,$rad. The X~mode (not shown) is degenerate with the A~mode for $\theta=0$ and is asymptotic to the O~mode for $ \theta \neq 0 $. As the plasma becomes relativistic, that is decreasing from $\rho\gg1$ to $\rho\ll1$, the dispersion curves become highly elongated very close to the light line.  (From RMM1.)}
\label{fig:RPE1}  
\end{figure}

\subsection{Three wave modes}

The wave properties at radio frequencies, $\omega\ll\Omega_{\rm e}$, in ${\cal K}$ can be summarized as follows. There are three modes. One of these is the X~mode which has vacuum-like dispersive properties for $\beta_{\rm A}^2\gg1$ and a polarization that precludes it being generated through a resonant beam-driven instability. The other two modes are referred to here as the L~and A~modes for parallel propagation, and as the O~and Alfv\'en modes for oblique propagation. A conventional way of plotting a dispersion relation is frequency as a function of wavenumber, that is $\omega$ as a function of $k_\parallel$ in the present case. Dispersion curves are shown on such a plot in Figure~\ref{fig:RPE1} for a case where the spread in energies is nonrelativistic, specifically for a 1D J{\"u}ttner distribution $\propto e^{-\rho\gamma}$ with $\rho=20$, corresponding to a temperature $T=mc^2/\rho\approx3\times10^8\,$K. The solid curve and solid (diagonal) line are the dispersion relations for $\theta=0$, corresponding to the L~and A~modes, respectively. The X~mode is degenerate with the A~mode for $\theta=0$ and is asymptotic to the O~mode for $ \theta \neq 0 $.  The L~mode curve in Figure~\ref{fig:RPE1} may be interpreted as a plot of $\Re [z^2\Re W(z)]^{1/2}$ versus $1/z$. 

It is convenient to choose the independent variable to be $z=\omega/k_\parallel c$, rather than $k_\parallel$. The dispersion relations are
{\br 
\be
\omega=\omega_L(z)=[\omega_p^2z^2\Re W(z)]^{1/2},
\qquad
z=z_A,
\label{pdrs}
\ee
}
for the L~mode and $z=z_A$ for the A~mode.  Figure~\ref{fig:RPE1} is plotted for a value of $\beta_{\rm A}\gg1$ such that the line $z=z_A$ cannot be distinguished from the light line $z=1$.

For slightly oblique propagation, the two modes reconnect to form the O~mode and the Alfv\'en mode. The reconnection occurs at $z=z_A$, $\omega=\omega_{\rm co}$, where
{\br
\be\omega_{\rm co}=\omega_L(z_A)
\label{omegaco}
\ee
}
is referred to as the cross-over frequency. The dispersion curve for the nearly parallel ($\theta\to0$) O~mode is $\omega\approx\omega_L(z)$ for $z>z_A$ and $z\approx z_A$ for $\omega>\omega_{\rm co}$, and the dispersion curve for the  the nearly parallel Alfv\'en is $z\approx z_A$ for $\omega<\omega_{\rm co}$ and for $\omega\approx \omega_L(z)$ for $z<z_A$. For nonzero $\theta$ the frequency of the oblique modes is given by \citep[][RMM1]{MG99}
\be
\omega^2(z,\theta)=\frac{\omega_L^2(z)}{1+a(z)\tan^2\theta},
\qquad
a(z) = \frac{b}{z_A^2 - z^2},
\label{omega_oblique}
\ee
with $ b \approx 1 $ for $ \beta_{\rm A} \gg 1 $. Near $\omega_{\rm co}$, $a(z)$ is very large in magnitude, and the two dispersion curves move away from each other very rapidly with increasing $\theta\ll1$: the O~mode moves to higher $\omega$ and larger $z$ and the Alfv\'en mode moves to lower $\omega$ and smaller $z$ with increasing $\theta$. The condition $ \omega^2(z,\theta) \geq 0 $ implies $ z^2 > z_A^2 + b \tan^2\theta $ for the O~mode and $ z^2 < z_A^2 $ for the Alfv\'en mode. The O~mode is superluminal for $ \theta $ satisfying $ z_A^2 + b\tan^2\theta > 1 $ which may be approximated as $  \theta \gtrsim 1/\beta_{\rm A} $.

\begin{figure}
\centering
 \psfragfig[width=1.0\columnwidth]{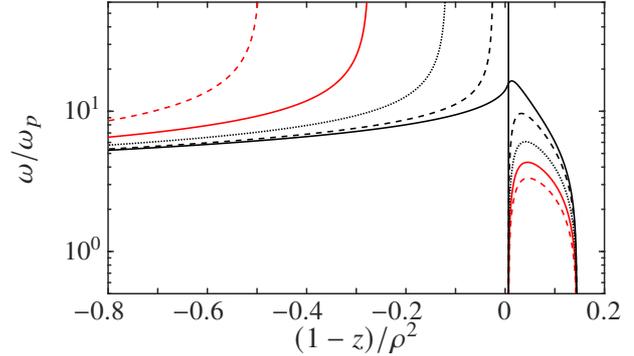}
\caption{Dispersion curves in a highly relativistic case, $ \rho = 0.01 $ ($\av{\gamma}\approx100$), $ \beta_{\rm A} \approx 1.2\times 10^3$ (corresponding to $ r/r_{\rm LC} \approx 0.122 $) and $ \theta = 0 $ (black solid), $0.25\rho\,$rad (black dashed), $0.5\rho\,$rad (black dotted), $0.75\rho\,$rad (red solid), and $0.1\rho\,$rad (red dashed). The black solid curve corresponds to the L~mode, the solid vertical line at $ z = z_A $ corresponds to the A~mode with the O~mode to its upper left and the Alfv\'en mode to its lower right. The Alfv\'en mode exists between $z = z_A$, which is very close to $1-z=0$ in the figure with $ \gamma_A = 8.7\times10^2$, and  $z=z_0$. The maximum in the dispersion curve occurs near $z=z_m$. (From RMM1.)}
\label{fig:RPE2}  
\end{figure}

\subsection{Effect of increasing $\av{\gamma}$}

The dispersion curves in Figure~\ref{fig:RPE1} are for a nonrelativistic spread, $\av{\gamma}-1\ll1$, and they are strongly modified by relativistic effects for $\av{\gamma}\approx1/\rho\gg1$. We compare the dispersion curve for the L~mode in the nonrelativistic and highly relativistic cases for both superluminal ($1-z<0$) and subluminal ($1-z>0$) phase speeds. In the superluminal region, for the nonrelativistic case shown in Figure~\ref{fig:RPE1}, the  cutoff frequency, $\omega_x$, corresponding to $z\to\infty$, is slightly below $\omega_{\rm p}$  and the frequency, $\omega_1$, at which the dispersion curve crosses the light line, $z=1$, is slightly above $\omega_{\rm p}$. 

For $\av{\gamma}\gg1$, in the superluminal range, the frequency increases with decreasing $z$ from $\omega_x=\omega_{\rm p}\langle\gamma^{-3}\rangle^{1/2} \approx \omega_{\rm p}/\av{\gamma}^{1/2}$ at $z=\infty$ to $\omega_1\approx\omega_{\rm p}(2\av{\gamma})^{1/2}$ at $z=1$. In the subluminal region, $\omega$ initially increases with decreasing $z$, as is evident in Figure~\ref{fig:RPE1}, with the frequency of the L~mode reaching a maximum, at $z=z_m$, and then decreasing to zero at $z=z_0$ along a second branch. In this range the frequency is a double-valued function of $k_\parallel$ (cf. Figure~\ref{fig:RPE1}), which we refer to as the upper-$z$ and lower-$z$ branches. The waves in the higher-$z$ branch have some similarities with Langmuir waves, in that they are longitudinal and subluminal. However, these waves are unlike Langmuir in other ways: they exist only for $\gamma_\phi>\gamma_m\gg1$, the ratio of the electric energy to the total energy in the waves is very small ($\approx1/24\av{\gamma}^2$ rather than $\approx1/2$ for Langmuir waves) and their group speed is very close to unity. The lower-$z$ branch corresponds to negative dispersion and strong Landau damping; we do not discuss such waves here.

For subluminal $z$ in the relativistic case, the parameter $ a(z) $ in~(\ref{omega_oblique}) may be approximate  by
\be
a(z)\approx\frac{\beta_{\rm A}^2\gamma_\phi^2}{\beta_{\rm A}^2-\gamma_\phi^2},
\label{az}
\ee
where we assume $ \beta_{\rm A} \gg 1 $, $\gamma_\phi\gg1$. The O~mode is subluminal only for a tiny range of $z\lesssim1$ corresponding to $\gamma_\phi\gtrsim\beta_{\rm A}/(1-\beta_{\rm A}^2\theta^2)^{1/2}$ and hence $a(z)<0$. The Alfv\'en mode has $\gamma_\phi\lesssim\beta_{\rm A}$ and hence $a(z)>0$, with $a(z)$ changing sign by passing through infinity at $\gamma_\phi=\beta_{\rm A}$.

A (linear) plot of $\omega$ vs $k_\parallel$ (or $[z^2\Re W(z)]^{1/2}$ vs $1/z$) is not convenient for illustrating the dispersive properties in the subluminal range $1-z\ll1$ for $ \langle \gamma \rangle\approx1/\rho \gg 1 $. An alternative plot shown in  Figure~\ref{fig:RPE2} is of the logarithm of $\omega/\omega_{\rm p}$ against $(1-z)/\rho^2$ (RMM1) with $ z = 1 $ corresponding to $(1-z)/\rho^2 = 0 $, $ z> 1 $ to its left and $ z < 1 $ to its right. Near $z=1$ the form of the dispersion relation scales in a simple way with $\av{\gamma}\approx1/\rho$, such that a plot of $(\omega/\omega_{\rm p})/\av{\gamma}^{1/2}$ versus $(1-z)\av{\gamma}^2$ is approximately independent of $\av{\gamma}\gg1$. The peak is $z_m^2\Re W(z_m)\approx2.7\av{\gamma}$ at $z=z_m\approx 1 - 0.013/\av{\gamma}^2$, or $\gamma_\phi=\gamma_m\approx6\av{\gamma}$. The region of negative dispersion and strong Landau damping is $z_0<z\lesssim z_m$, with $z_0\approx1-0.14/\av{\gamma}^2$, or $\gamma_0<\gamma_\phi\lesssim\gamma_m$ with $\gamma_0\approx2\av{\gamma}$. 

The cross-over frequency, $\omega_{\rm co}=\omega_L(z_A)$, is between the peak in the RPDF at $z=z_m$ and the light line $z=1$ for $z_A>z_m$, corresponding to $\beta_{\rm A}>\gamma_m\approx6\av{\gamma}$ for $ \beta_{\rm A} \gg 1$.  We assume this inequality to be satisfied. If this were not the case, either the cross-over is in the region of negative dispersion, $z_0<z=z_A<z_m$, or at  $z=z_A<z_0$, when the two dispersion curves do not cross. The region $z<z_m$ is to the right of the maximum in the curves in Figure~\ref{fig:RPE2}. We do not discuss waves in the region of negative dispersion, $z<z_m$ ($\gamma_\phi<6\av{\gamma}$), assuming them to be too heavily damped to be of relevance.

The maximum frequency of the Alfv\'en mode is a function of $\theta$, as shown in Figure~\ref{fig:RPE2}. This maximum frequency is
\be
\omega_{A{\rm max}}(\theta)\approx \frac{1.7\,\omega_{\rm p}\av{\gamma}^{1/2}}{(1+\gamma_\phi^2\theta^2)^{1/2}}\approx\frac{1.7\,\omega_{\rm p}\av{\gamma}^{1/2}}{\gamma_\phi\theta},
\label{omegaAmax}
\ee
where we assume $\gamma_\phi^2\ll\beta_{\rm A}^2$, and $\gamma_\phi\theta\gg1$ in the latter approximation. Waves near this maximum, although on the Alfv\'en branch, are quite different from conventional Alfv\'en waves; we refer to them as being on the ``turnover'' branch.\footnote{There is another intrinsically oblique, low-frequency mode, that corresponds to $ \omega_L^2(z) < 0 $, cf. RMM1, that we do not discuss here.}

The properties of these two modes for $\av{\gamma}\gg1$ may be summarized as follows. The O~mode exists for $\omega\ge\omega_x=\omega_{\rm p}\langle\gamma^{-3}\rangle^{1/2}\approx\omega_{\rm p}/\av{\gamma}^{1/2}$  and is superluminal except for a tiny range of angles, $\theta\lesssim1/\beta_{\rm A}$, at $\omega>\omega_1\approx\omega_{\rm p}(2\av{\gamma})^{1/2}$. The Alfv\'en mode has its conventional dispersion relation, written here as $z=z_A$, with $z_A\approx1-1/2\beta_{\rm A}^2$ for $\beta_{\rm A}^2\gg1$, only at sufficiently low frequencies; the dispersion curve deviates to smaller $z$ with increasing frequency, with a maximum frequency at $z=z_m$, and with this maximum  decreasing $\propto1/\theta$ with increasing $\theta \gg 1/\gamma_\phi$.

\subsection{Subluminal waves}

Both the Cerenkov and anomalous Doppler resonances require that the resonant waves be subluminal. There are weakly damped subluminal waves only for $\gamma_\phi\gg6\av{\gamma}$. 

Subluminal O~mode waves have $\gamma_\phi>\beta_{\rm A}\gg1$ and $\theta<1/\beta_{\rm A}$. Beam-driven wave growth of O~mode waves is possible in principle only for this tiny range of angles, $\theta\lesssim1/\beta_{\rm A} \approx 2\times10^{-4}$~rad for $ \langle \gamma\rangle \approx 10 $ at $ r/r_{\rm LC} = 0.1 $. The resonance condition $\gamma_\phi = \beta_{\rm b} $ requires a beam with $\gamma_{\rm b}>\beta_{\rm A}$ to resonate with O~mode waves at $\theta\to0$, increasing to $\gamma_{\rm b}\gg\beta_{\rm A}$ as $\theta$ increases towards $1/\beta_{\rm A}$. These conditions apply in the rest frame ${\cal K}$ of the plasma, and in the pulsar frame ${\cal K}'$ an additional factor $2\gamma_{\rm s}$ appears, for example, $\theta\lesssim1/\beta_{\rm A}$ becomes $\theta'\lesssim1/2\gamma_{\rm s}\beta_{\rm A}$ and $\gamma_\phi>\beta_{\rm A}$ becomes $\gamma'_\phi>2\gamma_{\rm s}\beta_{\rm A}$. 

The Alfv\'en mode is always subluminal. In ${\cal K}$ its dispersion relation is well approximated by $\gamma_\phi=\beta_{\rm A}$ at low frequencies, with $\gamma_\phi<\beta_{\rm A}$ at higher frequencies, as the maximum frequency (\ref{omegaAmax}) is approached for $\gamma_\phi\approx6\av{\gamma}$.  In ${\cal K}'$ these become $\gamma'_\phi=2\gamma_{\rm s}\beta_{\rm A}$ at low frequencies, with $\gamma'_\phi<2\gamma_{\rm s}\beta_{\rm A}$ at higher frequencies and the maximum frequency at $\gamma'_\phi\approx6\gamma_{\rm s}\av{\gamma}$.

\section{Critique of CCE}
\label{sect:CCE}

The major difficulty with CCE is the coherence mechanism. We first summarize the problem of coherent emission from a more general perspective, identifying three forms of coherent emission: reactive (or hydrodynamic or self-bunching) instabilities, kinetic (or maser) instabilities and superradiance. We then discuss application of these to CCE.

\subsection{Coherence mechanisms}
\label{sect:coherence}

\citet{GZ75} classified coherence mechanisms as maser or antenna mechanisms. A maser mechanism is well-defined: it involves negative absorption. In its simplest form an antenna mechanism involves a bunch of $N$ particles radiating $N^2$ times the power emitted spontaneously by one particle. We separate antenna mechanisms into two classes, which we refer to as reactive instabilities and superradiance, depending on how the bunch is formed. In a reactive instability the emission process itself (here curvature emission) causes self-bunching, and feedback from the bunching causes the amplitude of the wave to grow. In most discussions of CCE, the existence of the bunch is either postulated as an initial condition or is attributed to some physical process unrelated to curvature emission, such as soliton formation. It is this form of coherence that we refer to as superradiance, which may be described as an enhanced (by constructive interference) form of spontaneous emission. Although superradiance was originally defined by \citet{D54} in terms of an initial array of quantum oscillators, classical counterparts are well known \citep{Aetal80,GH82}. In models for CCE the superradiance is attributed to such phase-coherent spontaneous emission associated with solitons or other structures. 

\subsection{Self-bunching and CCE}

In a reactive instability there is feedback between the wave field and particle bunching such that the two grow in unison. An early suggestion for self-bunching due to curvature emission  \citep{GK71} was based on an idealized model of relativistic particles moving around a ring. This suggestion stimulated some early critical discussion \citep{Saggion75,CR77,BB78}. A related instability was proposed by \citet{BGI87,BGI88a}, and this also led to criticism \citep{LP87} and controversy \citep{BGI88r,LP88}.

The acceleration that causes curvature emission in a magnetic field is due to the Lorentz force associated with the curvature drift velocity \citep{CES75}. This velocity is of magnitude $v_c=\gamma\beta^2 c^2/R_c\Omega_{\rm e}$ and is directed across the field lines. One may attribute a self-bunching instability associated with curvature emission to the curvature drift. The curvature-drift instability was discussed by \citet{KL10}, who estimated the growth factor and concluded that it is too small. These authors argued that this self-bunching instability should be excluded from the list of potential mechanisms for pulsar radio emission. Following \citet{KL10} we conclude that the self-bunching form of CCE is not viable for pulsars.

\subsection{Maser CCE}
\label{sect:maser}

For a distribution of relativistic particles in 1D motion along a circular path, the absorption coefficient corresponding to curvature emission is similar in form to that for synchrotron emission. Synchrotron absorption can be negative only under special conditions, and the same applies to curvature absorption \citep{B75,M78,ZS79,CS88,LM92,LM95}. While maser curvature emission is possible in principle, the growth rate is too small for it to be relevant for pulsars, as the following remarks indicate.

Maser curvature emission is driven by a positive gradient of the distribution function (summed over electrons and positrons) with respect to energy. This is the same driver as for the maser (or kinetic) form of the beam-driven instability of L-mode waves. In RMM3 we found that the reactive version of the weak-beam instability does not exist for a J{\"u}ttner distribution. Although we have not explored whether or not the reactive version of the curvature-drift instability exists for a J{\"u}ttner distribution, we argue on general grounds that reactive growth (when it exists) is faster than kinetic growth. The argument \citep{KL10} that reactive growth is too slow to be effective applies {\it a fortiori}\/ to maser growth. We conclude that the maser form of CCE is also not viable for pulsars.

\subsection{Superradiance in CCE}

Once self-bunching and maser curvature emission are excluded, the remaining possible form of the coherence required for CCE to operate is some form of bunching caused by a mechanism that is not related to curvature emission. We identify such a mechanism as a classical version of superradiance. The idea is that when individual charges are arranged in an initial configuration, the spontaneous emission from these charges can occur in phase. In an ideal case this leads to $N$ charges radiating $N^2$ times the power in spontaneous emission per single charge.

As an aside we remark on a notable qualitative difference between superradiance and the other two forms of coherent emission concerning the sign of the charge. There is no superradiance in a pair plasma if the electron and positron distributions are identical; this is because the radiative electric fields due to the positive and negative charges cancel. However, for the other two forms of coherent emission, the contributions of electrons and positrons to the absorption coefficient or the growth rate have the same sign, for both beam-driven and curvature-driven maser and reactive instabilities. To be effective, the postulated bunching mechanism in CCE must lead to a bunch with a net charge.

We interpret as superradiance the coherence mechanism postulated in models for CCE developed in the 1970s \citep{R69,K70,S71,RS75,BB76a,BB77b,CR77}. Criticism of this form of CCE \citep{Kirk80,M81}, led to some early controversy \citep{BB83}, and resulted in an ongoing diversity of views between supporters and critics of CCE. The ongoing diversity of views concerns the viability or otherwise of the suggested mechanism (soliton formation) for the bunching. 

\subsection{Soliton-based CCE}

In a soliton-based model for CCE, the soliton formation is assumed (implicitly) to restore the putative initial configuration continuously such that the coherent emission is continuous.
The suggestion that the coherence is due to bunches associated with solitons involves two instabilities: a beam-driven instability to generate waves, usually assumed to be Langmuir-like waves, and a modulational instability that leads to these waves forming solitons. An early version of this suggested mechanism was discussed critically by \citet{Karpman_etal75}, and these authors came to a negative conclusion concerning the possibility of explaining pulsar radio emission in terms of coherent curvature radiation resulting from soliton formation. Later authors  \citep[e.g.,][]{Buti78,MP80,MP84,A93,MGP00,Mitra17,LMM18} argued that the soliton formation should occur. 

Our primary argument against soliton-based CCE is that resonant beam-driven growth is ineffective in a pulsar plasma, implying that the growth (required to produce the Langmuir-like waves) does not occur. The argument that this is the case is discussed below in connection with RPE. Suppose we ignore this argument and assume that resonant beam-driven growth were effective, as a first stage in RPE. One could then regard CCE as one of several possibilities for the second stage. Other possibilities are induced scattering, LAE and FEM, as discussed further below. In this context, soliton formation leading to CCE is just one of several competing second-stages processes, any of which could potentially lead to escaping radiation. However, none is relevant if beam-driven growth of Langmuir-like waves is ineffective. 

Even if a soliton does form, it needs to be charged in order to produce coherent emission. The suggested modulational instability, described by the nonlinear Schr\"odinger equation \citep[e.g.,][]{MGP00,LMM18}, causes bunching through the ponderomotive force, which does not depend on the sign of the charge. To form a charged soliton it is assumed \citep{MGP00} that the electrons and positrons have different mean Lorentz factors, ${\bar\gamma}_\pm={\bar\gamma}\pm\Delta\gamma/2$ say, such that the relative motion between them results in the current density ($\propto\Delta\gamma/{\bar\gamma}^3$) required by pulsar electrodynamics. The different mean Lorentz factors imply that the electrons and positrons respond differently to the ponderomotive force, resulting in a charge separation within the soliton \citep{MGP00}.  This is a very small effect,
but it is required for the soliton to have a net charge.

\subsection{Is any form of CCE viable for pulsars?}

Despite CCE being widely favored (primarily for observational reasons) as the pulsar radio emission mechanism, the (theoretical) arguments against it seem compelling. Self-bunching and maser instabilities for curvature emission are possible in principle, but fail quantitatively. Most important, the assumed beam-driven growth of Langmuir-like waves, required as the first stage in the assumed soliton formation, does not occur in a pulsar plasma that is intrinsically relativistic in the sense $\av{\gamma}-1\gtrsim1$. We conclude that CCE based on beam-driven wave growth and resulting soliton formation is not plausible as the pulsar radio emission mechanism.

\section{Critique of beam-driven RPE}
\label{sect:RPE}

There are severe constraints on beam-driven RPE in a pulsar plasma with $\av{\gamma} - 1 \gtrsim 1$. Pre-conditions for growth are the resonance condition, which requires $\gamma_\phi \lesssim \gamma_{\rm b} $, wave dispersion, which requires $\gamma_\phi\gtrsim6\av{\gamma}$ (RMM1), and the separation condition for the beam and background, which requires $\gamma_{\rm b}\gtrsim 10\av{\gamma}$ (RMM2) for a weak-beam system where $ n_{\rm b}/\gamma_{\rm b}n_0 \ll 1 $.\footnote{For equal counter-streaming distributions the separation condition is $\gamma_{\rm b}\gg2\av{\gamma}^2$ (RMM2).} The maximum growth rate occurs when $ \gamma_\phi \approx (10{\rm-}20)\av{\gamma} \approx \gamma_{\rm b, min} $ (RMM3). The possible growth rates and the inhomogeneous structure of the pulsar plasma lead to further constraints.

\subsection{Beam-driven nearly-parallel waves}

The largest growth rate for a beam-driven instability is for parallel propagation in the L~mode. With $\gamma_\phi \approx \gamma_{\rm b, min}$ from the resonance condition, for $\beta_{\rm A}<\gamma_{\rm b}<\infty$ the L~mode approximates the O~mode for propagation angle $ \theta \ll 1 $, and for $\gamma_{\rm b}<\beta_{\rm A}$ the L~mode approximates the Alfv\'en mode for $ \theta \ll 1 $, cf. Figure~\ref{fig:RPE1}, Figure~\ref{fig:RPE2} and RMM1. In discussing the magnitude of the growth rate it is not important to distinguish between these two cases.

For slightly oblique O~mode waves the resonance condition requires $\gamma_\phi>\beta_{\rm A}/(1 - \beta_{\rm A}^2\theta^2)^{1/2}$, with there being no subluminal O~mode waves for $\theta\gtrsim1/\beta_{\rm A}$. The condition $\gamma_{\rm b}>\beta_{\rm A}/(1 - \beta_{\rm A}^2\theta^2)^{1/2}$ is not plausibly satisfied for the parameters estimated in Section~\ref{sect:parameters}. The estimate $\beta_{\rm A}\geq5.1\times10^3$ at $r/r_{\rm LC}\leq0.1$ for $ \av{\gamma} \approx 10$, based on equation (\ref{parameters}), requires a beam with $\gamma_{\rm b}\geq5.1\times10^3$, increasing $\propto(r/r_{\rm LC})^{-3/2}$ for a source at lower heights $r/r_{\rm LC}<0.1$. These numbers are not compatible with the multiple-beam model discussed above for a bulk streaming speed $\gamma_{\rm s}=10^2$--$10^3$ for the background which would require the beam to have bulk streaming Lorentz factor $\sim 10^6 {\rm -} 10^7 $. The resonance condition can be satisfied for the O~mode only if one assumes a beam with a much higher Lorentz factor than the  multiple-beam model allows for plausible values.

\subsection{Beam-driven Alfv\'en mode}

The resonance condition is less restrictive for oblique Alfv\'en waves than for O~mode waves. The dispersion curve for the Alfv\'en mode may be separated into three portions, as shown in Figure~\ref{fig:RPE2}: a low-frequency Alfv\'en-like portion with dispersion relation $z\approx z_A$ or $\gamma_\phi\approx\beta_{\rm A}$, a turnover portion in the range $\beta_{\rm A}>\gamma_\phi>6\av{\gamma}$, near the maximum frequency given by equation (\ref{omegaAmax}), and a negative-dispersion portion where the frequency decreases with decreasing $z$ or $\gamma_\phi$. The threshold condition $\gamma_{\rm b} \gtrsim \gamma_\phi\gtrsim 10\av{\gamma}$ implies that only part of the turnover portion is relevant; the negative-dispersion portion is of no relevance.

RPE based on beam-driven Alfv\'en waves, on the Alfv\'en-like portion, has been suggested as a possible pulsar emission mechanism \citep{TK72,Letal82,MG99,L00}. Resonance on the Alfv\'en-like portion of the dispersion curve requires $\gamma_{\rm b}\approx\beta_{\rm A}$. In a slowly varying magnetosphere, with $\beta_{\rm A}\propto1/r^{3/2}$, this condition can be satisfied at only one particular height $r$ for a given $\gamma_{\rm b}$. As discussed above in connection with the O~mode, the condition $\gamma_{\rm b}\approx\beta_{\rm A}$ cannot be satisfied for plausible parameters. Our estimate below is for the growth rate on the turnover portion.

\subsection{Growth rate}

It is convenient to introduce the fractional growth rate, $\Gamma/\omega$, where $\Gamma$ is the e-folding rate of growth of wave energy. The maximum growth rate is for the L~mode, for which we approximate the dispersion relation by $\omega=\omega_L(z)\approx\omega_L(1)\approx\omega_{\rm p}(2\av{\gamma})^{1/2}$. We (RMM3) estimated the maximum fractional growth rate  in ${\cal K}$ for the kinetic weak-beam instability, with equal $\av{\gamma}$ for the beam and the background, finding
\be
\frac{\Gamma}{\omega}\approx\left(\frac{n_{\rm b}}{\gamma_{\rm b}n_0}\right)^{1/2}\frac{1}{2\av{\gamma}^3},
\label{rri}
\ee 
where $ n_{\rm b} $ is the number density of the beam in the rest frame of the background, with $\omega\approx\omega_{\rm p}(2\av{\gamma})^{1/2}$.

In RMM3 we discussed the relation between temporal and spatial growth rates, and their transformation between inertial frames. In brief, given the temporal growth rate $\Gamma$ in the rest frame ${\cal K}$, the spatial growth rate is $\beta_{\rm g}\Gamma/c$ in ${\cal K}$. In ${\cal K}'$ the temporal and spatial growth rates are $(\gamma'_{\rm g}/\gamma_{\rm g})\Gamma$ and $(\gamma'_{\rm g}\beta'_{\rm g}/c\gamma_{\rm g})\Gamma$, respectively, with $\beta'_{\rm g}=(\beta_{\rm g}+\beta_{\rm s})/(1+\beta_{\rm g}\beta_{\rm s})$ the group speed in ${\cal K}'$, and with $\gamma_{\rm g}$, $ \gamma'_{\rm g} = \gamma_{\rm s}\gamma_{\rm g}(1 + \beta_{\rm g}\beta_s) $ the Lorentz factors corresponding to the group speeds in ${\cal K}$, ${\cal K}'$, respectively. The group speed in ${\cal K}$ is $\beta_{\rm g}=z[1-2R_{\rm L}(z)]$ (RMM1), and for $\gamma_\phi^2=\gamma_{\rm b}^2\gg1/4R_{\rm L}(z)$, one has $\gamma_{\rm g}\approx2.5\av{\gamma}$, where we make the approximation $R_{\rm L}(z)\approx R_{\rm L}(1)\approx1/24\av{\gamma}^2$.

\subsection{Outward and inward growing waves in ${\cal K}$}

In a multi-beam model, there are assumed to be clouds with different bulk speeds, and in this case we identify the frame ${\cal K}$ as that in which the mean bulk speed is zero. In ${\cal K}$ individual clouds are assumed to have a range of bulk speeds with positive (outward) and negative (inward) values, and the relative speed of one cloud overtaking another can be either positive or negative. In a given overtaking event, resonant waves with $z \lesssim \beta_{\rm b}$ are either outward ($z>0$, $\beta_{\rm g}>0$) or inward ($z<0$, $\beta_{\rm g}<0$) in ${\cal K}$. (We only consider positive energy waves: $ z $ and $ \beta_{\rm g} $ has the same sign or $ z\beta_{\rm g} > 0 $. We have $ \beta_{\rm g} (z) = z [1 - 2 R_{\rm L}(z)] $ with maximum value of $ R_{\rm L}(z) = 1/2 $ at $ z = \infty $ (RMM1). Over the range of interest, $ \gamma_\phi > \gamma_m $, we have $ z\beta_{\rm g} > 0 $ always.) The inward propagating waves in ${\cal K}$ are outward propagating in ${\cal K}'$ for $\{|z|,|\beta_{\rm g}|\} < \beta_{\rm s}$, so that both outward and inward propagating waves  in ${\cal K}$ are potential candidates for pulsar radio emission, which is assumed to be propagating outward in ${\cal K}'$. It is convenient to label these two cases as $\pm$, and to compare these for given $z=\pm|z|$, $\beta_{\rm g}=\pm|\beta_{\rm g}|$. 

Assuming that the frequencies of the waves are the same in ${\cal K}$, $\omega_\pm=\omega$, and that the resonance condition is satisfied, $z=\pm\beta_{\rm b}$, the frequencies in ${\cal K}'$ are 
\be
\omega'_\pm=\gamma_{\rm s}\omega(1 \pm \beta_{\rm s}/\beta_{\rm b}),
\label{omegpm}
\ee
with $ \omega_- < 0 $ for $ \beta_{\rm s} > \beta_{\rm b} $ or $ \gamma_{\rm s} > \gamma_{\rm b} $. Note that $\omega'_-$ is negative and that the interpretation is based on the dispersion equation being unchanged under $\omega',k'_\parallel \to -\omega',-k'_\parallel $, such that the negative-frequency backward-propagating wave is re-interpreted as a positive-frequency forward-propagating wave in ${\cal K}'$. For $ \{\gamma_{\rm s}^2, \gamma_{\rm b}^2, \gamma_{\rm g}^2\} \gg 1 $ we may write
\be
\omega'_+\approx2\gamma_{\rm s}\omega, 
\quad
\omega'_-\approx-(\gamma_{\rm s}/2\gamma_{\rm b}^2)\omega,
\label{omegpm_approx}
\ee
where $\gamma_{\rm s}^2\gg\gamma_{\rm b}^2$ is assumed in the latter case. For a pulsar plasma we have $ \gamma_{\rm s} = 10^2{\rm-}10^3 $ with maximum growth rate when $ \gamma_{\rm b} = \gamma_\phi \approx (10{\rm -}20)\av{\gamma} $ (RMM3). We use $ \gamma_{\rm b} \approx 15\av{\gamma} $ henceforth. At resonance we have $ \gamma_{\rm b} = \gamma_\phi $ so that $ \gamma_{\rm s}^2 \gg \gamma_{\rm b}^2 $ is satisfied in general for $ 2\lesssim \av{\gamma} \lesssim 10 $ or $ 1 \gtrsim \rho \gtrsim 0.1 $ which we consider as relevant to pulsars. 

\subsection{Fractional growth rates in ${\cal K}'$}

The growth rates in ${\cal K}'$ for the $\pm$-cases are related to those in ${\cal K}$ by the ratios of the group speeds in the two frames. In ${\cal K}'$, the group speeds and Lorentz factors are
\be
\beta'_{\rm g\pm}=\frac{\pm\beta_{\rm g}+\beta_{\rm s}}{1\pm\beta_{\rm g}\beta_{\rm s}},
\qquad
\gamma'_{\rm g\pm}=\gamma_{\rm g}\gamma_{\rm s}(1\pm\beta_{\rm g}\beta_{\rm s}),
\label{betagp}
\ee
giving $\gamma'_{\rm g+}\approx2\gamma_{\rm g}\gamma_{\rm s}$ and $\gamma'_{\rm g-} \approx {\rm max}\{\gamma_{\rm s}/2\gamma_{\rm g}, \gamma_{\rm g}/2\gamma_{\rm s}\}$. Assuming the same fractional growth rates in ${\cal K}$, the fractional growth rates in ${\cal K}'$ follow from (\ref{omegpm}) and (\ref{betagp}). These give
\be
\left(\frac{\Gamma'}{\omega'}\right)_\pm  \!=\! \frac{1 \pm \beta_{\rm g}\beta_{\rm s}}{1 \pm \beta_{\rm s}/\beta_{\rm b}}\left(\frac{\Gamma}{\omega}\right),
\quad
\left(\frac{\Gamma'}{\omega'}\right)_+ \!\approx\left(\frac{\Gamma}{\omega}\right),
\quad
\left(\frac{\Gamma'}{\omega'}\right)_- \!\approx \frac{\gamma_{\rm b}^2}{\gamma_{\rm g}^2}\left(\frac{\Gamma}{\omega}\right),
\label{fgrs}
\ee
where in the final expression we assume $ \gamma_{\rm s}^2 \gg \{\gamma_{\rm g}^2, \gamma_{\rm b}^2\} $, with $\omega'_-$ assumed positive, as discussed above. For $ \gamma_{\rm g} \approx 2.5\av{\gamma} $ and $ \gamma_{\rm b} \approx 15\av{\gamma} $ we have $ \gamma_{\rm b}^2/\gamma_{\rm g}^2 \approx 40 $.

\subsection{Frequency of growing waves in ${\cal K}'$}

For growth to result in waves in the frequency range observed, the frequency, $\omega=\omega_\pm$, of the waves in ${\cal K}$ must transform into a frequency, $\omega'=\omega'_\pm$, in ${\cal K}'$ that is in the observed range of pulsar radio emission.

\begin{figure}
\centering
\psfragfig[width=1.0\columnwidth]{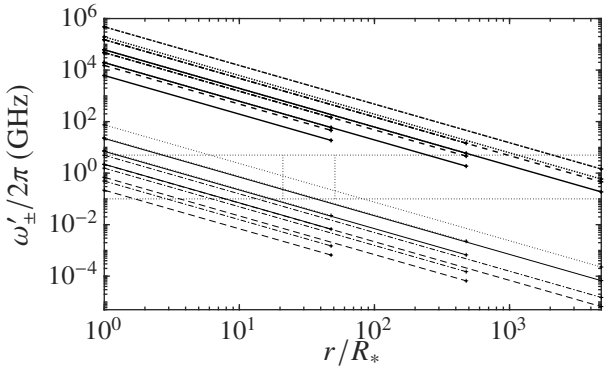}
\caption{Plots of $ \omega'_+/2\pi $ (thick) and $ \omega'_-/2\pi $ (thin) for $ P = 0.1, 1, 10\, {\rm s} $ corresponding to lines extending to $ r/R_* = 0.1 r_{\rm LC}/R_* \approx 48, 4.8\times10^2, 4.8\times10^3 $, respectively, terminated by a marker; $ \gamma_{\rm s} = 10^2 $ (solid and dashed), $ 10^3 $ (dotted and dash-dotted); $ \rho = 1 $ (solid and dotted), 0.1 (dashed and dash-dotted). We use $ \kappa = 10^5 $ and $ \dot{P}/P^3 = 10^{-15}\,{\rm s}^{-3} $ for all plots. The thin dotted horizontal lines are at $ 0.1 $ and 5 GHz and the two thin dotted vertical lines are at $ r/R_* = 20 $ and 50.}
\label{fig:omega_prime_pm}  
\end{figure}

\begin{figure}
\centering
\psfragfig[width=1.0\columnwidth]{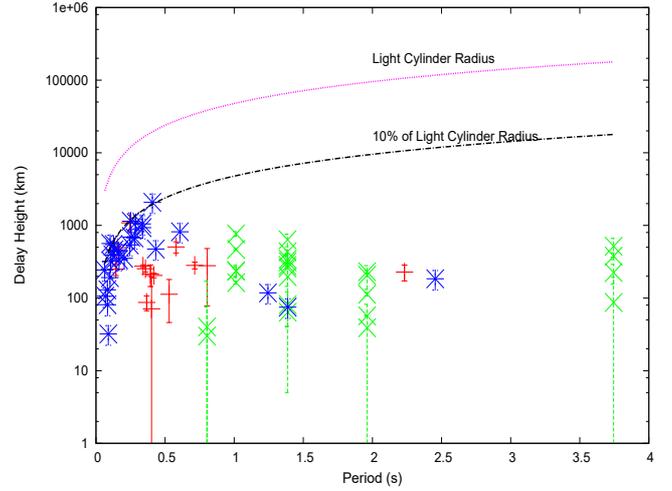}
\caption{Plots of emission height (above the stellar surface) as a function of pulsar period \citep[from][]{Mitra17}. The blue point at emission height of 32\,km (or $ r/R_* \approx 4 $) is the Vela pulsar with a period of 89.33\,ms \citep{WJ08}.}
\label{fig:Mitra_2017}  
\end{figure}

The frequencies $\omega_\pm=\omega$ of the waves in ${\cal K}$ may be approximated by $\omega=\omega_{\rm p}(2\av{\gamma})^{1/2}$. In Figure~\ref{fig:omega_prime_pm} we show plots of $ \omega'_+/2\pi $ (thick) and $ \omega'_-/2\pi $ (thin) for $ P = 0.1, 1, 10\, {\rm s} $ corresponding to lines extending to $ r/R_* = 0.1 r_{\rm LC}/R_* \approx 48, 4.8\times10^2, 4.8\times10^3 $, respectively, terminated by a marker; $ \gamma_{\rm s} = 10^2 $ (solid and dashed), $ 10^3 $ (dotted and dash-dotted); and $ \rho = 1 $ (solid and dotted), 0.1 (dashed and dash-dotted). We use $ \kappa = 10^5 $ and $ \dot{P}/P^3 = 10^{-15}\,{\rm s}^{-3} $ for all plots. The thin dotted horizontal lines are at $ 0.1 $ and 5 GHz indicating the frequency range of pulsar radio emission. We see that $ \omega'_+/2\pi $ is far too large except near $ r/r_{\rm LC} = 0.1 $ for slowly rotating pulsars; and $ \omega'_-/2\pi $ is generally too small except very close to the pulsar surface. \cite{Mitra17} estimated the emission height between $ r/R_* = 20 $ and 50 (indicated by thin dotted vertical lines) for pulsars regardless of their period. It is evident that neither of $ \omega'_\pm $ cover the range of radio frequencies from pulsars for plausible parameter values. However, the range of emission heights given by \cite{Mitra17} is an average and as can be seen from Figure~\ref{fig:Mitra_2017} the emission height estimates range from about $ r/R_* = 4 $ (Vela pulsar PSR J0835-4510 with a period of $ 89.33\,{\rm ms} $) to about 200 (PSR J0835-4510 with a period of $ 0.41\, {\rm s} $) \citep{WJ08}. Furthermore, while the emission height of $ r/r_{\rm LC} = 0.1 $ appears to be an upper limit for fast rotating pulsars, the emission height of slower pulsars are much smaller than this upper limit.
In discussing the second requirement, we ignore these complications,
and allow the emission height to extend from the stellar surface to $ r/r_{\rm LC} = 0.1 $.

\subsection{Possible propagation paths}

For a wave to propagate in an inhomogeneous, time-independent plasma it must be directed along a path on which its frequency remains constant. The maximum growth rate is for parallel propagation. Assuming a path parallel to the magnetic field, the wave frequency $\omega=\omega_{\rm p}[z^2W(z)]^{1/2}$, must remain constant as $\omega_{\rm p}$ decreases along this path in order for the wave to escape, implying that the RPDF $z^2W(z)$ must increase along this path. The RPDF is a decreasing function of increasing $z$ over the range $z_m<z<1$. It follows that for escaping waves $z$ must decrease along the escape path, and this decrease is limited by $z>z_m$. 

Positive wave growth at $z$ requires that the distribution function, $g(u)$, be an increasing function of $u$ or $\beta$ at $z=\beta$. For a beam with speed $\beta_{\rm b}$, growth occurs over a range $z<\beta_{\rm b}$. In a weak-beam model the separation condition (RMM2) requires that the total distribution function has a minimum between the background distribution and the beam. For an escaping wave, wave growth turns to damping when $z=\beta$ reaches this minimum. The bandwidth of the growing waves may be interpreted as the range of frequencies corresponding to $z=\beta$ in the range between this minimum and $\beta_{\rm b}$. 

This conclusion is not modified significantly by considering oblique propagation. In the oblique case the dispersion relation is given by (\ref{omega_oblique}), with the dependence on angle described by the denominator which is positive and an increasing function of angle for $\gamma_\phi<\beta_{\rm A}$. In principle, the frequency of an escaping wave that is initially oblique may remain constant due to $\theta$ decreasing, so that this denominator decreases. However, the maximum growth rate is for $\theta=0$, so that if growth does occur it greatly favors waves with small obliquity, $\theta\ll1$. The scope for allowing the wave frequency to remain constant due to decreasing obliquity is very restricted, and we ignore this possibility.

\subsection{Bandwidth of growing waves}

For wave growth to be effective the growth factor must be large, e.g., $G\gtrsim30$, where $G$ is the number of e-folding wave growths. This factor is estimated as the spatial growth rate times the distance over which a given wave grows. Assuming parallel propagation, this distance is identified as that over which the resonant frequency changes, due to the change in $\omega_{\rm p}$ with distance, by the bandwidth, $\Delta\omega$ say, of the growing waves. 

In the nonrelativistic case, $\Delta\omega$ is estimated from the spread, $\Delta\beta$ say, over which the growth rate is near its maximum value, corresponding to a range $\Delta z=\Delta\beta$ of phase speed such that the bandwidth is $k_\parallel c\Delta\beta$. The estimate of $\Delta\omega$ for a RSG distribution is a straightforward generalization of the nonrelativistic case. However, this is not the case for the distributions considered here. In an intrinsically relativistic plasma, the dispersion relation of the relevant waves, $\omega=\omega_L(z)$, is a rapidly varying function of $z$, and the estimate of the bandwidth of the growing waves needs to take this into account.

The bandwidth of waves in a small range $\Delta z$ may be estimated as $\Delta\omega=\Delta z d\omega/dz$, with $\omega^2=\omega_{\rm p}^2z^2W(z)$. An approximation is to estimate the derivative of the RPDF at $z=1$ for a J\"uttner distribution, as in RMM1. This gives $\Delta\omega/\omega\approx 12\av{\gamma}^2\Delta z$. The range $\Delta z$ of speed corresponds to a range $\Delta\gamma_\phi=\gamma_\phi^3\Delta z$ of Lorentz factor. This leads to the estimate of the fractional bandwidth in ${\cal K}$
\be
\frac{\Delta\omega}{\omega}\approx\frac{\Delta\gamma_\phi}{\gamma_\phi^3}12\av{\gamma}^2.
\label{bandwidth1}
\ee

In the following discussion we do not attempt to estimate the fractional bandwidth in detail, but leave $\Delta\omega/\omega$ as a parameter of order unity. The rationale for this is that that $\gamma_\phi$ in (\ref{bandwidth1}) is less than but order $\gamma_{\rm b}$, and the spread $\Delta\gamma_\phi$ is of the same order. Also the value $12\av{\gamma}^2$, made for $z=1$ is an underestimate, due to the slope of $z^2W(z)$ increasing for $z<1$, until it starts to decrease as $z$ approaches $z_m$. One expects $\gamma_{\rm b}$ to be of order several times $\av{\gamma}$, resulting in the right hand side of (\ref{bandwidth1}) being of order unity.

We compare inward and outward propagating waves in ${\cal K}$ generated by otherwise identical beams propagating in opposite direction, so that the waves have the same frequency, $\omega=\omega_\pm$, in ${\cal K}$. The frequencies in ${\cal K}'$ are given approximately by (\ref{omegpm_approx}). The bandwidth, $\Delta\omega_\pm=\Delta\omega$ in ${\cal K}$ are also the same. The frequencies and bandwidths transform in the same way, such that the fractional bandwidths $\Delta\omega'_\pm/\omega'_\pm$ are equal to the fractional bandwidth, $\Delta\omega/\omega$ in ${\cal K}$. The frequency of beam-generated waves in ${\cal K}$ is in the range $2.7\av{\gamma}\gtrsim\omega^2/\omega_{\rm p}^2>2\av{\gamma}$, between the peak in the RPDF at $z=z_m$ and the frequency at $z=1$. An approximate estimate of the frequency is $\omega=1.5\omega_{\rm p}\av{\gamma}^{1/2}$. The peak in the RPDF corresponds is at $\omega/\omega_{\rm p}\approx1.64$, giving an estimate of the fractional bandwidth $\Delta\omega/\omega\lesssim0.1$, with $\Delta\omega/\omega\ll0.1$ if the separation between the minimum and maximum $\beta$ in $g(u)$ is sufficiently small.

\subsection{Growth factor for weak-beam model}

Growth of a wave at a given frequency occurs only over the distance $(\Delta\omega')_\pm L_\parallel/c\beta'_{g\pm}$ over which the resonant frequency remains within the bandwidth of the growing waves. This implies that the growth factor, $G_\pm$, is given by
\be
G_\pm=\frac{\Gamma'_\pm\Delta\omega'_\pm}{\omega'_\pm} \frac{L_\parallel} {c\beta'_{g\pm}}
=\frac{\Gamma'_\pm}{\omega'_\pm}\frac{\Delta\omega}{\omega}
{\omega'_\pm} \frac{L_\parallel} {c\beta'_{g\pm}},
\label{Gpm1}
\ee
where the same relative bandwidth applies to the $\pm$ cases in both frames. Using the relations (\ref{omegpm}), (\ref{betagp}) for $\gamma_{\rm s}\gg\gamma_{\rm g}$, (\ref{fgrs}), and inserting the expression (\ref{rri}) for the fractional growth rate, (\ref{Gpm1}) gives
\be
    G_+
        \approx \gamma_{\rm s} \left(\frac{n_{\rm b}}{\gamma_{\rm b}n_0}\right)^{1/2}\frac{1}{\av{\gamma}^3}
        \frac{\Delta\omega}{\omega}
        \omega\frac{L_\parallel}{c},
        \qquad
        G_-
        \approx\frac{G_+}{4\gamma_{\rm g}^2},
        \label{Gpm2}
\ee
where we assume $\beta'_{\rm g\pm}\approx1$. With $\omega\approx1.5\omega_{\rm p}\av{\gamma}^{1/2}$ and $\omega_{\rm p}\propto n_0^{1/2}$, the growth factor is independent of the density of the background plasma. An alternative way of writing  (\ref{Gpm2})  is
\be
    G_+
        \approx \frac{3\gamma_{\rm s} }{4\av{\gamma}^{5/2}}
        \frac{\Delta\omega}{\omega}
    \frac{\omega_{\rm b}L_\parallel}{c},
        \qquad
        G_-
        \approx\frac{G_+}{4\gamma_{\rm g}^2},
        \label{Gpm3}
\ee
with $\omega_{\rm b}=(e^2n_{\rm b}/\varepsilon_0m\gamma_{\rm b})^{1/2}$ interpreted as a plasma frequency corresponding to the beam. The estimate $\gamma_{\rm g}\approx2.5\av{\gamma}$ made above corresponds to $4\gamma_{\rm g}^2\approx25\av{\gamma}^2$. The growth factor is proportional to the bandwidth of the growing waves in ${\cal K}'$ and the fact that $\Delta\omega'_-$ is smaller than $\Delta\omega'_+$ explains why $G_-$ is smaller than $G_+$.

\subsection{Estimate of growth factors for a weak beam}

There is considerable uncertainty in using (\ref{Gpm3}) to estimate the growth factor for a weak-beam model: the parameters $\omega_{\rm b}$ and $L_\parallel$ are poorly determined, and there is a strong (implicit) dependence on the height of the source, which is also uncertain. Here we consider only order of magnitude estimates. 

Assuming $ n_{\rm b}/n_0 = 10^{-3} $  \citep{ELM83} and $\gamma_{\rm b}=10^2$, one has $\omega_{\rm b}\approx3\times10^{-3}\omega_{\rm p}$. For $ \kappa = 10^5 $, $ \dot{P}/P^3 = 10^{-15}\,{\rm s}^{-3} $, $\gamma_{\rm s}=10^3$, $P=1\,$s, (\ref{parameters}) gives $\omega_{\rm p}\approx10^7\rm\,s^{-1}$ at $r/r_{\rm LC}=0.1$ and $\omega_{\rm p}\approx2\times10^{10}\rm\,s^{-1}$ at $r/R_*=30$. The factor $\gamma_{\rm s}/\av{\gamma}^{5/2}$  is of order unity for $\av{\gamma}=10$. For $\Delta\omega/\omega\approx0.1$, these estimates give $G_+$ of order $0.1\omega_{\rm b}L_\parallel/c$ and $G_-$ smaller than $G_+$ by a factor $25\av{\gamma}^2\approx3\times10^3$ for $\av{\gamma}\approx10$. One has $0.1\omega_{\rm b}L_\parallel/c\approx10^{-6}L_\parallel$ at $r/r_{\rm LC}=0.1$ and  $0.1\omega_{\rm b}L_\parallel/c\approx10^{-3}L_\parallel$ at $r/R_*=30$, with $L_\parallel$ in meters. It is apparent from these rough estimates that effective growth, that is, $G_\pm\gtrsim30$, requires  very large $L_\parallel$ for $G_+$ and much larger values $L_\parallel$ for $G_-$. For example, the most favorable of these cases  for effective growth is for $G_+$ at $r/R_*=30$ where $L_\parallel\gtrsim30\,$km would be required. In contrast, effective growth for $G_-$ at $r=30R_*=3\times10^4\,$km would require $L_\parallel\gtrsim10^5\,$km.

In a smoothly-varying model for the magnetosphere, the plasma frequency varies $\propto r^{-3/2}$, implying a characteristic length $L_\parallel=3r/2$ for changes in $\omega_{\rm p}$. We conclude that even for this estimate of $L_\parallel$, effective growth seems marginally possible for the $+$~case for a source relatively close to the stellar surface, and is not possible for the $-$~case. Any local inhomogeneities imply smaller $L_\parallel$, giving a more restrictive limit on the growth factor. Specifically, in a multi-beam model, $L_\parallel$ depends on the length of and separation between individual clouds of pairs, so that $L_\parallel\ll3r/2$ is expected. We conclude that even when growth is assumed to be possible, effective growth requires larger $L_\parallel$ and $\Delta\omega/\omega$ than is plausible even in the most favorable cases.

An important proviso is that this negative conclusion relies on our assumption that the background plasma is intrinsically relativistic, with our numerical estimated sensitive to our assumed value $\av{\gamma}\approx10$. For smaller $\av{\gamma}-1\gtrsim1$ the numerical constraints on the growth factor are weaker. 

\subsection{Second stage of RPE}
\label{sect:second_stage}

Our definition of RPE implies two different physical processes operating to produce the emitted radiation. The first stage is assumed here, as in other discussions of RPE, to be a beam-driven instability. Several seemingly overwhelming difficulties are implied by the foregoing discussion of beam-driven wave growth. Nevertheless let us ignore these difficulties and suppose that subluminal L~waves do grow in an allowed narrow range of $z$ corresponding to $\gamma_\phi\gtrsim6\av{\gamma}$. These waves are on the turnover branch of the Alfv\'en mode for $\gamma_\phi<\beta_{\rm A}$ and on the O~mode branch for $\gamma_\phi>\beta_{\rm A}$.  We separate possibilities for the second stage into ``passive''  and ``active'' mechanisms.

\subsubsection*{Passive conversion}

In conventional plasma emission, the energy in Langmuir waves produced in the first stage is converted into escaping radiation passively, through nonlinear processes in the plasma or due to mode coupling through inhomogeneities in the plasma. Similarly, in principle, nonlinear processes  or inhomogeneities can lead to partial conversion of wave energy from a mode that cannot escape into waves in a mode that can escape \citep[e.g.,][]{I88,Lyubarsky96,U00,U02}. Such processes are ``passive'' in the sense that the total energy in waves is not changed during the conversion process. An interesting possibility, that does not exist for conventional plasma emission, is that the waves generated through the beam instability can escape directly. This applies to the O~mode (but not to the Alfv\'en mode), the dispersion curve for which has a subluminal range (for $\gamma_\phi>\beta_{\rm A}$, $\theta<1/\beta_{\rm A}$) that joins on to a superluminal range that corresponds to waves that can escape to infinity (to zero $\omega_{\rm p}$). However, the requirements on beam-driven O-mode growth are particularly severe, e.g., a beam with Lorentz factor $\gamma_{\rm b}>\beta_{\rm A}$ which is of order $3\times10^5$ according to (\ref{parameters}) for $r=30R_*$. For a more plausible $\gamma_{\rm b}$, just above the separation threshold $10\av{\gamma} \approx 100$ for $\av{\gamma}=10$, the waves are in the Alfv\'en mode.

Suppose that Alfv\'en waves on the turnover branch are generated in the pulsar plasma, and that a similar passive conversion occurs, with $\omega$ not changing significantly (apart from possible frequency doubling). Alfv\'en waves near the turnover frequency, 1.7$\omega_{\rm p}\av{\gamma}^{1/2}/\gamma_{\rm b}\theta$, can be converted into waves in other modes only if these modes have nearly the same frequency as the Alfv\'en waves. The two possibilities are O~mode and X~mode waves. Conversion into O~mode waves is possible only if the frequency of the Alfv\'en waves exceeds $\omega_x=\omega_{\rm p}\langle\gamma^{-3}\rangle^{1/2}$, which imposes the relatively weak constraint $\theta<1.7\av{\gamma}^{1/2}/\langle\gamma^{-3}\rangle^{1/2}\gamma_{\rm b} \approx 1.7\av{\gamma}/\gamma_{\rm b}$. There is no constraint on passive conversion to X~mode waves. We conclude that passive conversion into either mode is allowed kinematically.

\subsubsection*{Active conversion}

There are several suggested pulsar emission mechanisms (LAE, FEM, CCE) that rely on a large-amplitude wave (LAW) in the background plasma, and one possible way such waves can be generated is through a beam-driven instability. In an ``active'' second-stage process, the energy in the escaping radiation is attributed to the radiating particles through a process that relies on the presence of a LAW, including LAE, FEM and a favored form of CCE. In LAE (or FEM) it is assumed that the first stage of RPE results in a large-amplitude wave and the radiation is due to the accelerated motion of particles in the field of the wave. In CCE the radiation is due to the accelerated motion of particles along a curved field line, and the large-amplitude wave is invoked to provide the necessary coherence. We comment on LAE and FEM here and CCE is discussed in Section~\ref{sect:CCE}. A related second-stage mechanism is induced scattering, which is a passive process in a nonrelativistic plasma, but is an active process in a pulsar plasma where the frequency of the  waves scattered by the relativistic particles is much higher than the frequency of the LAW. Emission due to induced scattering in a pulsar plasma has been discussed by \citet{Lyubarsky96} and \citet{LP96}.  

In LAE \citep{C73,M78,R92a,R92b,R95,ML09,MRL09,RK10}, the basic emission process is that due to the accelerated (1D) motion of a charge in a parallel electric field. The absorption coefficient for LAE, due to a distribution of relativistic particles, can be negative, causing maser-like emission.  One needs to distinguish between LAWs that are subluminal and superluminal. For a subluminal wave there exists a frame, moving with the phase velocity of the wave, in which the oscillations are purely spatial; in this frame the emission may be interpreted as FEM rather than LAE. In FEM \citep{FK04,FEM02} the acceleration by the field of the wave (or ``wiggler'') may be perpendicular or parallel to the direction of motion of the particle. For a superluminal wave there exists a frame, moving at the inverse of the phase speed, in which the oscillations are purely temporal, and LAE may be attributed to acceleration by the electric vector of the wave in this frame. In a model in which the LAW is attributed to beam-driven growth, the LAW is necessarily subluminal.

The characteristic frequency of LAE due to particles accelerated (periodically) to $\gamma$ in a wave of frequency $\omega_0$ is $\omega_0\gamma^2$. Assuming a beam-driven LAW propagating outward in ${\cal K}$, its frequency (for plausible parameters) is higher that the observed range for pulsars, and the extra boost by $\gamma^2$ exacerbates this problem. This problem with excessively high frequency is alleviated by assuming that LAE is due to a beam-driven LAW propagating inward in ${\cal K}$. 

These active second-stage processes, along with CCE, are relevant only if the first-stage mechanism can provide the necessary LAW. Our primary criticism of these suggested mechanisms applies to the assumed beam-driven wave growth, which we argue might occur under special conditions but cannot be the emission mechanism for all pulsars.

\section{Critique of ADE}
\label{sect:ADE}

The suggestion that ADE is the pulsar radio emission mechanism \citep[e.g.,][]{MU79,KMM91a,LBM99} must overcome the difficulty that the natural frequency of such emission is too high. The anomalous Doppler resonance condition is given by equation (\ref{gyroresonance}) with $s=-1$. Whereas as the resonance for $s=0$ requires $z=\beta$, the resonance for $s=-1$ requires $z<\beta$; we write these requirements as $\gamma_\phi=\gamma$ and $\gamma_\phi<\gamma$, respectively. The resonance condition for ADE then becomes 
\be
\omega=\frac{2\gamma_\phi^2\gamma^2}{\gamma^2-\gamma_\phi^2}\frac{\Omega_{\rm e}}{\gamma}
\approx2\gamma_\phi^2\frac{\Omega_{\rm e}}{\gamma},
\label{ADE1}
\ee
where the approximation applies for $\gamma^2\gg\gamma_\phi^2$. Assuming that all particles are in their ground (Landau) state,  the anomalous Doppler transition to the first excited state can, in principle, drive wave growth for all values of $\gamma$ for which equation (\ref{ADE1}) is satisfied. It follows that values of $\gamma$ near where the distribution function is maximum are favored, provided that equation (\ref{ADE1}) is satisfied.

The major difficulty with ADE as the pulsar radio emission mechanism is that the frequency (\ref{ADE1}) is too high. The frequency $\omega$ also needs to satisfy a dispersion relation, for one of the X, O and Alfv\'en modes. For nearly parallel propagation the dispersion relations  for the X~and Alfv\'en modes may be approximated by $z=z_A$ or $\gamma_\phi=\beta_{\rm A}$. The approximate form of equation (\ref{ADE1}) then requires $\omega/\Omega_{\rm e}\approx2\beta_{\rm A}^2/\gamma$. For any plausible location of the source of the radio emission one requires $\omega/\Omega_{\rm e}\ll1$ and hence $\gamma\gg2\beta_{\rm A}^2$. For the estimates made in equation (\ref{parameters}), this implies that ADE in these modes would be in the radio range only for impossibly high values of $\gamma$.

The lowest frequency consistent with equation (\ref{ADE1}), for given $\Omega_{\rm e}/\gamma$, is for the smallest value of $\gamma_\phi^2$. The only weakly damped waves that can exist in a pulsar plasma have $\gamma_\phi\gg\gamma_m\approx6\av{\gamma}$ in ${\cal K}$. Setting $\gamma_\phi\approx6\av{\gamma}$ in equation (\ref{ADE1}) gives $\omega/\Omega_{\rm e}=72\av{\gamma}^2/\gamma$ in ${\cal K}$. In the pulsar frame one has $\omega'\approx2\gamma_{\rm s}\omega$ and $\gamma'\approx2\gamma_{\rm s}\gamma$, so that this condition becomes $\omega'/\Omega_{\rm e}=(4\gamma_{\rm s})^2(72\av{\gamma}^2/\gamma')$ in ${\cal K}'$. This also implies that for ADE, at the lowest allowed value of $\gamma_\phi$, the frequency of emission would be in the radio range only for impossibly high values of $\gamma$.

We conclude that the intrinsic frequency (\ref{ADE1}) of ADE is too high to account for pulsar radio emission anywhere inside the light cylinder for the fiducial parameters chosen in (\ref{parameters}). 

\section{Discussion}
\label{sect:discussion}

Our objective in this paper is to explore the suggested pulsar radio emission mechanisms critically, to determine whether any of them is viable as a generic pulsar radio emission mechanism. We find that all suggested mechanisms encounter major difficulties, some of which are well known, and others are associated with the formation of relativistic beams and with the properties of wave dispersion in a pulsar plasma with $\av{\gamma}-1\gtrsim1$. Here we summarize our arguments in three categories: assumptions about the distributions of particles, assumptions about the wave dispersion and estimates of the wave growth. We also make specific comments about CCE, RPE and ADE.

\subsubsection*{Pair plasma}

It is widely accepted that pulsar plasma, in the source region of the radio emission, is dominated by pairs generated in cascades through one-photon pair creation. We make the following points.
\begin{itemize}

\item Models for the pair creation \citep[e.g.,][]{HA01,AE02} suggest an intrinsically relativistic spread in energies, $\av{\gamma}$ between a few and about ten in its rest frame ${\cal K}$ of the plasma.
Implicit or explicit assumptions that the plasma is cold or nonrelativistic, $\av{\gamma}-1\ll1$ in ${\cal K}$ can be seriously misleading concerning the wave properties, compared with $\av{\gamma}-1\gtrsim1$.

\item We argue that the default choice for the distribution with $\av{\gamma}-1\gtrsim1$ should be a J{\"u}ttner distribution. However, the properties of the wave dispersion are sensitive to $\av{\gamma}$ but not to the form of the distribution \citep{MG99}. 

\item Models for the pair creation also suggest a highly relativistic bulk streaming speed, e.g., $\gamma_{\rm s}$ of order $10^2$--$10^3$ in the pulsar frame ${\cal K}'$. We argue that such streaming should be included by applying a Lorentz transformation to the distribution function in ${\cal K}$.

\item The widely favored choice of a relativistic streaming Gaussian distribution (\ref{RSG}) is incompatible with any Lorentz-transformed distribution. Any model for a streaming distribution obtained by Lorentz transforming a given distribution in the rest frame is much broader than a Gaussian distribution for $\gamma_{\rm s}\gg1$. 

\item The separation condition, for a beam distribution with bulk Lorentz factor $\gamma_{\rm b}$ to be separated in momentum space from a background distribution at rest, is $\gamma_{\rm b}>2\av{\gamma}^2$ for equal number densities and $ \gamma_{\rm b} \gtrsim 10\av{\gamma} $ for a weak beam (RMM2).

\end{itemize}

\subsubsection*{Formation of beams}

The formation of the beams invoked to explain wave growth through a beam-driven instability is usually attributed to non-steady pair creation leading to separate clouds of pairs producing a beam when faster particles from a trailing cloud over take slower particles in a leading cloud. In \S\ref{sect:multiple-sparking} we discuss one specific model for such overtaking and argue that it encounters an overwhelming difficulty: the overtaking time is so long that the cloud has propagated out of the magnetosphere before any beam forms. In Appendix~\ref{app:convection} we develop a kinetic-theory model to describe this effect and associated fractionization, which is usually ignored, raising other potential difficulties with beam formation in such multiple-cloud models. Despite these difficulty seemingly ruling out any form of beam-driven wave growth, we ignore it and proceed to discuss possible wave growth, simply postulating that appropriate beams exist. 

\subsubsection*{Wave properties}

The properties of wave dispersion in a pulsar plasma are not taken into account in some models for the radio emission. We make the following points (RMM1). \begin{itemize}

\item For $\av{\gamma}-1\gtrsim1$ and $\beta_{\rm A}\gg1$ there are no waves with nonrelativistic phase speed $z\ll1$ (or group speed $\beta_{\rm g}\ll1$) in ${\cal K}$.

\item For parallel propagation, subluminal, weakly-damped L-mode waves exist for $\gamma_\phi\gtrsim6\av{\gamma}$, A~mode and X~mode waves have $z=z_A=\beta_{\rm A}/(1+\beta_{\rm A}^2)^{1/2}<1$; superluminal L~mode waves exist for all $z>1$.  For oblique propagation, the L~mode separates into the O~mode for $z^2 \gtrsim z_A^2 + \tan^2\theta$, which is purely superluminal for $\theta>1/\beta_{\rm A}$, and the Alfv\'en mode for $z<z_A$, which has a ``turnover'' branch for $z_A\gtrsim z\gtrsim z_m$.

\item Resonant beam-driven wave growth is possible in principle for a beam with $\gamma_{\rm b}\gtrsim6\av{\gamma}$, allowing O~mode waves (with $\theta\ll1/\beta_{\rm A}$) to grow for $\gamma_{\rm b}>\beta_{\rm A}$ and Alfv\'en waves (on the turnover branch) to grow for $6\av{\gamma}\lesssim\gamma_{\rm b}<\beta_{\rm A}$.

\end{itemize}

\subsubsection*{Beam-driven wave growth}

For further discussion, we assume that wave growth occurs, and consider the growth rate in the most favored case of parallel-propagating L~mode waves (actually on the turnover branch of the Alfv\'en mode). We find the following.
\begin{itemize}

\item Beam-driven wave growth due to (a) primary particles with $\gamma$ of order $10^6$--$10^7$ propagating through the secondary pair plasma, and (b) relative motion between electrons and positrons in the secondary plasma, are ineffective, due to the growth rate being too small. 

\item In the widely favored multiple-cloud model, including the older ``multiple-sparking'' model, in which pair creation occurs in localized, transient bursts, resulting in clouds of pair plasma,  beam-driven growth is attributed to faster particles in a trailing cloud overtaking slower particles in a leading. Our critical examination of this model (\S\ref{sect:multiple-sparking}) implies that the conditions required for this multiple-beam model to lead to significant wave growth are not plausibly satisfied. We argue that this class of model is ineffective in generating beams that cause wave growth.

\item In a multi-cloud model, growth of both outward ($+$~case) and inward ($-$~case) propagating waves in ${\cal K}$ needs to be considered. On transforming to the pulsar frame ${\cal K}'$, in which both are propagating outward, the frequency of the former is too high for $ r/r_{\rm LC} < 0.1 $ and the frequency of the latter is too low unless very close to the stellar surface, compared with the observed range of pulsar radio emission.

\item Effective growth requires a growth factor (number of e-folding growths), $G_\pm\gg1$, for either of these cases. We estimate the growth factor in ${\cal K}'$ and find $G_+\gg G_-$, with only $G_+$ possibly satisfying this condition for a source at $r/R_*\lesssim30$. A proviso is that our estimates are made for $ \av{\gamma}=10$ and that both $G_\pm$ are larger for smaller values of $ \av{\gamma}$.

\item The growth factor is proportional to the scalelength, $L_\parallel$, of the gradient in the plasma frequency. In a smoothly-varying model for the magnetosphere one has $L_\parallel=3r/2$. A much smaller $L_\parallel$ applies to localized clouds, giving a much smaller estimate of the growth factor.

\end{itemize}

\subsubsection*{CCE, RPE and ADE}

Our critical assessment of the three favored radio emission mechanism may be summarized as follows.
\begin{itemize}

\item CCE: In principle, maser and reactive versions of curvature emission are possible, but are ineffective in practice. The favored model of coherence, due to soliton formation, relies on beam-driven wave growth, with a modulational instability assumed to lead to soliton formation. Our arguments that beam-driven growth is ineffective implies that this form of CCE does not occur.

\item RPE: The foregoing points lead to the conclusion that effective beam-driven wave growth is not possible in a pulsar plasma. This excludes RPE as the pulsar radio emission mechanism. The least unfavorable case for beam-driven wave growth is for inward-propagating (in ${\cal K}$) wave on the turnover branch of the Alfv\'en mode. If such waves were to grow, a second stage involving passive conversion into escaping radiation in the X~mode or the O~mode would imply emission at a frequency that is arguably too low except near pulsar surface. Large-amplitude versions of such growing waves would also be candidates for active conversion through LAE of FEM, and then the frequency of the emitted radiation is arguably too high.

\item ADE: As with beam-driven growth, the resonance condition for ADE requires subluminal waves, and the fact that such waves exist only for $\gamma_\phi\gtrsim6\av{\gamma}$, along with the estimates in (\ref{parameters}), implies that the frequency of ADE is too high to explain all pulsar radio emission.

\end{itemize}

Our negative conclusions concerning the viability of the versions of CCE, RPE and ADE discussed here as plausible generic pulsar radio emission mechanisms depend on a number of assumptions. It is possible that one (or more) of our assumptions is inappropriate and that changing it might allow effective beam-driven wave growth, contrary to what we find. We comment on some such possibilities.

First, it may be that the model we assume for the pulsar magnetosphere and the pulsar plasma in it is incorrect. We assume that the radio source is in the polar-cap region that is populated by relativistically outflowing pair plasma. As already remarked, it has recently been suggested that the radio source may be beyond the light cylinder \citep{2019ApJ...876L...6P,2019MNRAS.483.1731L}, in which case our arguments are not directly relevant. 

Second, we assume that the plasma is intrinsically relativistic, streaming at $\gamma_{\rm s}=10^2$--$10^3$ with a spread $\av{\gamma}\lesssim10$ in its rest frame. If the spread is not relativistic, $\av{\gamma}-1\ll1$, then our arguments based on there being no wave modes with nonrelativistic phase speed would change: beam-driven growth would be possible under easily satisfied conditions, such as for solar type~III radio bursts. However, this seems unlikely: the large spread is intrinsic to a pair plasma generated by pair cascades, and assuming $\av{\gamma}-1\ll1$ would involve abandoning the long-standing assumption that the plasma is generated by pair cascades \citep{S71}. Although the pair-cascade model is based primarily on theoretical considerations, it remains the basis for the interpretation of the emission at both gamma-ray and radio frequencies  \citep[e.g.,][]{Petal15,Petal16}; abandoning it for the radio emission is not a plausible option.

Third, we assume that beam formation is due to faster particles in a trailing cloud overtaking slower particles in a leading cloud, in a multiple-cloud model, and find that for $\av{\gamma}\lesssim10$ the overtaking takes an impossibly long time. This time is reduced by assuming $\av{\gamma}\gg10$, but this would be inconsistent with models for the pair creation. To avoid this negative conclusion some other assumption for effective beam formation is required.

Fourth, we argue that the choice of a streaming distribution should be based on Lorentz transforming a plausible rest-frame distribution, and that this greatly reduces, compared with a relativistically streaming Gaussian (RSG) distribution, the growth rate of instabilities. This constraint would be relaxed if there were a physical argument in favor of a RSG distribution.

Fifth, in estimating growth factors, $G_\pm$, for outward and inward propagating waves in ${\cal K}$, both of which are propagating outward in ${\cal K}'$, we assume that in both $\pm$ cases the wave frequency is the same $\omega\approx1.5\av{\gamma}^{1/2}\omega_{\rm p}$. We find that $G_-$ is too small to lead to significant growth, and that $G_+$ could imply effective growth. The frequency, $\omega'_+=2\gamma_{\rm s}\omega$ of the outward propagating wave is too high to be relevant for most radio emission. This high frequency could be reduced by arguing for a smaller value of $\omega_{\rm p}$, for example due to multiplicity $\kappa\ll10^5$, but this would reduce $G_+$ by the same factor, implying ineffective growth.

\section{Conclusions}
\label{sect:conclusions}

Our analysis of various aspects of models for beam formation and beam-driven wave growth leads us to conclude that they encounter overwhelming difficulties. In particular, this conclusion implies that none of the currently favored emission mechanisms is viable as the generic pulsar radio emission mechanisms. The suggested mechanism CCE and RPE are based on beam formation and transfer of energy through a beam-driven plasma instability that generates subluminal waves. It is these basic processes that we find not to be viable.  An alternative emission mechanism that involves superluminal waves would avoid these difficulties. We discuss a model based on this alternative in an accompanying paper.

\section*{Acknowledgements}

The research reported in this paper was supported by the Australian Research Council through grant DP160102932. We thank an anonymous referee for helpful comments on the manuscript.

\section*{Data Availability}
No new data were generated or analysed in support of this research.

\bibliographystyle{mnras}

\bibliography{Pulsar_radio_Refs}

\appendix

\bsp	% typesetting comment
\label{lastpage}

\section{Distribution propagation}
\label{app:convection}

\begin{figure}
    \centering
    \tikzset{grid_lines/.style={very thin, color=blue!50}}
    \begin{tikzpicture}[scale=1]
        % trailing
        \draw (0,0) -- (0,1) -- (3,1) -- (3,0) -- cycle;
        % leading
        \draw (4,0) -- (4,1) -- (7,1) -- (7,0) -- cycle;
        % broadening
        \filldraw[fill=gray!20!white] (7,0) -- (7, 1) -- (7.5,1) -- (7.5,0) -- cycle;
        
        % labels
        \draw[|<->|] (0,-0.25) -- (3,-0.25) node[below,midway] {$ L'_0 $};
        \draw[|<->|] (3,-0.25) -- (4,-0.25) node[below,midway] {$ h'_0 $};
        \draw[|<->|] (4,-0.25) -- (7.5,-0.25) node[below,midway] {$ L'_0 + \Delta x'(t'_{02}) $};
        \node at (1.5,0.5) {$ f'_2(t'_{02}, x', u') $};
        \node at (5.75,0.5) {$ f'_1(t'_{02}, x', u') $};
        \draw[->] (1.5,1.25) -- (2.5,1.25) node[above, midway] {$ \gamma_2, \beta_2 $};
        \draw[->] (5.75,1.25) -- (6.75,1.25) node[above, midway] {$ \gamma_1, \beta_1 $};
        \draw (0,-0.75) -- (0,1.5);
        \node[rotate=90] at (-.25,0.375) {Stellar surface};
    \end{tikzpicture}
    \caption{Schematic diagram of the trailing cloud $ f'_2(t'_{02}, x', u') $, streaming velocity $ \beta_2 $, forming at the stellar surface after the tail of the leading cloud $ f'_1(t'_{02}, x', u') $, streaming velocity $ \beta_1 $, is a distance $ L'_0 + h'_0 $ from the stellar surface. The trailing cloud forms at time $ t' = t'_{02} $ during which the leading cloud has broadened by $ \Delta x'(t'_{02}) $. }
    \label{fig:g1g2_a}
\end{figure}
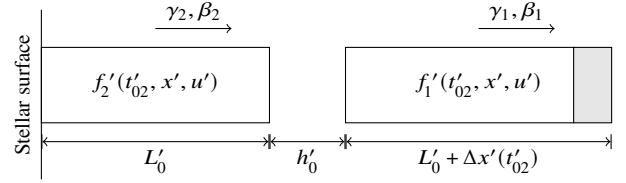

We assume that a leading cloud $ f'_1(t', x', u') $ of length $ L'_0 $ and (bulk) streaming Lorentz factor $ \gamma_1 $ is formed at the stellar surface. A trailing cloud $ f'_2(t', x', u') $ of length $ L'_0 $ and streaming Lorentz factor $ \gamma_2 $ is formed at the stellar surface once the slowest particle in the leading cloud is a distance $ L'_0 + h'_0 $ from the stellar surface. Figure~\ref{fig:g1g2_a} shows a schematic diagram of this scenario. The clouds lengthen as they propagate as discussed below.

We treat the propagation of the particle clouds using the 1D Vlasov equation in the absence of collision and external forces. In the pulsar frame we have, for $ i = 1, 2 $,
\begin{equation}\label{eq:Vlasov}
    \frac{\partial f_i'}{\partial t'} + c\beta'\frac{\partial f_i'}{\partial x'} = 0,
\end{equation}
where $ f_i' = f_i'(t', x', u') $ is the particle distribution, $ t' $ is the time, $ x' $ is the position, measured from the stellar surface, and $ u' = \gamma'\beta' $ is the particle 4-speed. We assume that
the plasma is initially distributed as
\begin{equation}\label{eq:f0}
    f'_i(t'_{0i}, x', u') = \left[H(x') - H(x' - L'_0)\right]g'_i(u'),
\end{equation}
where 
$ t'_{0i} $ is the time cloud $ i $ is created, $ g'_i(u') $ is the J\"uttner distribution and $ H(x') $ is the unit step function. 

The standard deviation of $ g_i(u) $ in the rest frame of the plasma, denoted as $ \sigma_{i} $, is given by $ \sigma_{i}^2 = \av{(u - \av{u}_i)^2}_i = \av{\gamma^2}_i - 1 $, where $ \av{Q}_i $ denotes the average of quantity $ Q $ over distribution $ g_i(u) $, with 68.23\%, 95.45\% and 99.73\% of the particles in distribution $ i $ within $ \sigma_i $, $ 2\sigma_i$ and $ 3\sigma_i $ of $ \av{u}_i = 0 $, respectively. Let $ u_{i\pm} = \pm 3\sigma_i = \pm3(\av{\gamma^2}_i - 1)^{1/2} $ then in the pulsar frame we have $ u'_{i\pm} = \gamma_{i\pm}\gamma_i (\beta_{i\pm} + \beta_i) $. We truncate the J\"uttner distribution $ i $ so that particles with $ u'_{i+} $ ($u'_{i-}$) are the fastest (slowest) particles of the cloud in the pulsar frame. The maximum travel distance available to particles within $ r/r_{\rm LC} \leq 0.1 $ is $ L_{\rm max} \sim 3R_*(1 {\rm -} 10^2) $, as seen in Figure~\ref{fig:Mitra_2017}, which corresponds to maximum allowed travel time $ t'_{i\pm,\rm max} \sim L_{\rm max}/c\beta'_{i\pm} $ for fastest and slowest particles, respectively, where
\begin{equation}\label{eq:betaipm_prime}
    \beta'_{i\pm} = \frac{\beta_{i} + \beta_{i\pm}}{1 + \beta_{i}\beta_{i\pm}}.
\end{equation}

Solving~\eqref{eq:f0} using method of characteristics gives $ f'_i(t', x', u') = f'_i(t'_{0i}, x' - c\beta' (t' - t'_{0i}), u') $, i.e. $ f'_i(t',x',u') $ remains unchanged along the characteristic curves $ x' = c\beta' (t' - t'_{0i}) + x'_{0i} $, where $ x'_{0i} = x'(t' = t'_{0i}) $. Thus we have
\begin{equation}
    f'_i(t', x', u')\! =\! \left[H(x' - c\beta' (t' - t'_{0i})) - H(x' - c\beta' (t' - t'_{0i}) - L'_0)\right]\! g'_i(u').
\end{equation}
We set $ t'_{01} = 0 $ which implies that $ t'_{02} = (L'_0 + h'_0)/c\beta'_{1-} $. In particular, two particles with velocities $ \beta'_1, \beta'_2 $ which are initially separated by $ \Delta x'_0 $ become separated by a distance
\begin{equation}
    \Delta x'(\Delta t') = c(\beta'_1 - \beta'_2) \Delta t' + \Delta x'_0,
\end{equation}
after a travel time of $ \Delta t' $.

\subsection{Fractionation}

\begin{figure}
\centering
\psfragfig[width=1.0\columnwidth]{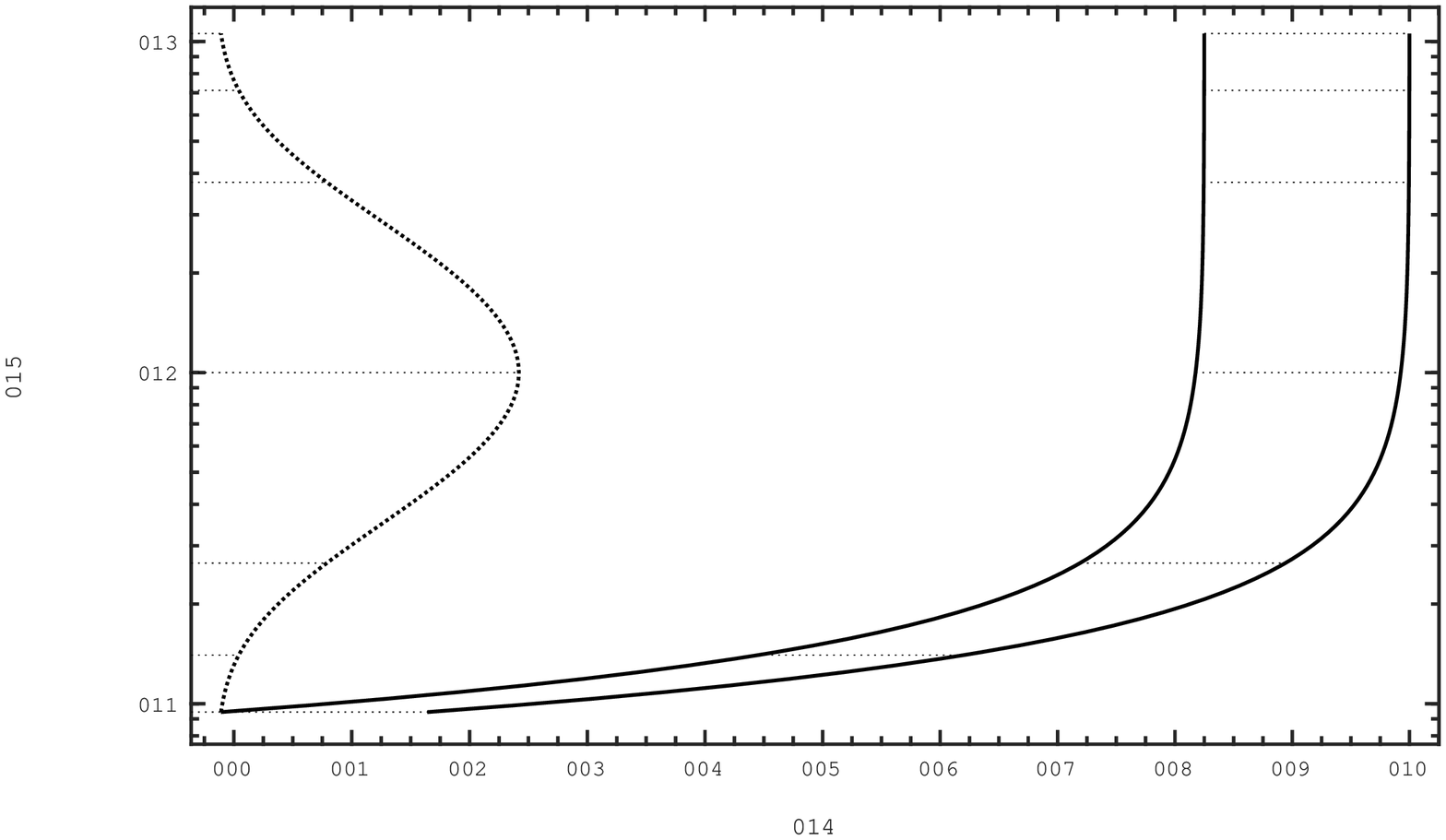}
\caption{Contour plots of $ f'_i(t', x', u') $ for $ \rho = 1 $, $ \gamma_i = 10^2 $ for a travel distance of $ 300R_* $. The two solid curves indicate the left and right edge of the plasma column for particles with 4-speed $ u' $ and the thin dotted lines joining them indicate contours lines at $ u = -3\sigma_{i} \ldots +3\sigma_{i}$. The height of $ f'_i(t',x',u') $ at these contour lines is shown schematically by the dotted curve on the left; which is the J\"uttner distribution $ g'_i(u') $.}
\label{fig:fractionation}  
\end{figure}

In an initially homogeneous plasma column propagating outward in the pulsar magnetosphere, faster particles travel a greater distance than slower particles during the same period of time. This effect is enhanced by longer travel time and greater spread in energy within the distribution. This results in fractionation of the plasma distribution. Significant levels of fractionation would invalidate the assumption of a homogeneous plasma.

Consider the slowest particles, $ \beta' = \beta'_{i-} $, and the fastest particles, $ \beta' = \beta'_{i+} $, in cloud $ i $ starting at the surface of the pulsar. In the time it takes the fastest particles to travel $ L_{\rm max} $, the slowest particle falls behind by $ \Delta x'(t'_{+,\rm max}) = (1 - \beta'_{i-}/\beta'_{i+}) L_{\rm max} \approx L_{\rm max}(\gamma'^2_{i+} - \gamma'^2_{i-})/2\gamma'^2_{i+}\gamma'^2_{i-} $, where the approximation applies for $ \gamma'_{i\pm} \gg 1 $. We caution that the assumption $ \gamma'_{i-} \gg 1 $ may not necessarily be satisfied when the spread in energy is large. Therefore, cloud $ i $ has a length of $ L'_0 + \Delta x'(t'_{+,\rm max}) $ by the time its fastest particles reach $ r/r_{\rm LC} = 0.1 $. \cite{AM98} stated that for pulsars one has $ h'_0 \sim 10^2 $\,m and $ L'_0 \sim (30{\rm -}40)h'_0 \sim (3{\rm -}4)\times10^3 $\,m. We choose nominal values of $ h'_0 = 10^2 $\,m and $ L'_0 = 3.5\times10^3 $\,m. Figure~\ref{fig:fractionation} shows contour plots of $ f'_i(t', x', u') $ for $ \rho = 1 $, $ \gamma_i = 10^2 $ for a travel distance of $ 300R_{*} $. The two solid curves indicate the left and right edge of the plasma column for particles with 4-speed $ u' $ and the thin dotted lines joining them indicate contours lines, from bottom to top, at $ u = -3\sigma_{i} \ldots +3\sigma $. The height of $ f'_i(t',x',u') $ at these contour lines is shown schematically by the dotted curve on the left; which is in fact the J\"uttner distribution $ g'_i(u') $. The initial length of cloud $ i $ is $ L'_0 = 0.35 R_{*} = 3.5\times10^3$\,m and the final cloud length is $ \approx L'_0 + 1.7R_{*} \approx 2\times10^4 $\,m corresponding to 5.7 times the original cloud length. The additional $ 1.7R_* $ length of the column is independent of the initial length $ L'_0 $ implying that the level of fractionation is greater for shorter plasma columns. As evident from Table~\ref{tab:Delta_L} the level of spatial lengthening of the distribution increases with both decreasing $ \gamma_i $ (i.e. longer travel time) and decreasing $ \rho $ (i.e. greater spread in energy).

\begin{table}
\begin{center}
    \bgroup
    \def\arraystretch{1.25}
    \begin{tabular}{llll}
    \toprule
    $ \gamma_i $          & $ \rho $  & $ \Delta x'(t'_{+,\rm max}) $ (m) & Percentage increase\\
    \midrule
    \multirow{2}{*}{$ 10^2 $}   & 1         & $ 1.7\times(10^2{\rm -}10^4) $    & $ 4.9\times(10^0{\rm -}10^2) $\\
                                & 0.1       & $ 9.3\times(10^3{\rm -}10^5) $    & $ 2.7\times(10^2{\rm -}10^4) $\\
    % \midrule
    \multirow{2}{*}{$ 10^3 $}   & 1         & $ 1.7\times(10^0{\rm -} 10^2) $   & $ 4.9\times(10^{-2}{\rm -}10^0) $\\
                                & 0.1       & $ 1.1\times(10^2{\rm -} 10^4) $   & $ 3.1\times(10^0{\rm -}10^2) $\\
    \bottomrule
    \end{tabular}
    \egroup
    \caption{Value of $ \Delta x'(t') $ for various values of $ \gamma_i $ and $ \rho $ at $ t' = t'_{+,\rm max} $. }
    \label{tab:Delta_L}
\end{center}
\end{table}

An important consequence of this spatial lengthening of the distribution is fractionation of an initially homogeneous cloud. For example, in the case shown in Figure~\ref{fig:fractionation} only particles with $ 0.12 < u'/u_i < 0.15 $ may be found at the center of the cloud. This effect is enhanced (diminished) for longer (shorter) travel distances and larger (smaller) spread of particle energies. The assumption that the plasma is homogeneous can become invalid as the cloud propagates in the pulsar magnetosphere for some parameter values.

\subsection{Cloud overlap}

In the pulsar frame, the background plasma is assumed to propagate with Lorentz factor $ \gamma_{\rm s} \sim 10^2{\rm -}10^3 $ \citep{AE02}. The separation condition implies that the beam must have $ \gamma_{\rm b} \gtrsim 10\av{\gamma} $ in the rest frame of the background \citep{RMM19b_2019JPlPh..85c9011R,RMM19_3}. This corresponds to the beam propagating with Lorentz factor $ \gamma_{\rm r} \approx 2\gamma_{\rm s}\gamma_{\rm b} \gtrsim 2\times(10^3{\rm-}10^4)\av{\gamma} $ in the pulsar frame. We identify the background with the slower leading cloud $ g'_1(u') $ and the beam with the faster trailing cloud $ g'_2(u') $ so that $ \gamma_1 \sim 10^2{\rm -}10^3 $ and $ \gamma_2 \gtrsim 20\gamma_1\av{\gamma} \gtrsim 4\times(10^3{\rm-}10^4) \gg \gamma_1 $.

In the rest frame of the background, one may have the beam travelling in either direction; only the separation in energy is important. In the pulsar frame, one may then freely assume that the beam is either the slower or the faster cloud. We assume the cloud is the faster of the two for the sake of simplicity. 

Whether the trailing cloud is able to overtake the leading cloud is subject to our definition of overlap between the two clouds. 

We consider two cases: (1) particles travelling with bulk outflow speed, $ \beta' = \beta_2 $ in the trailing cloud catch up to particles with bulk outflow speed, $ \beta' = \beta_1 $, in the leading cloud; and (2) the fastest particles in the trailing cloud with $ \beta' = \beta'_{2+}$ catch up to particles with bulk outflow speed $ \beta' = \beta_1 $ in the leading cloud.

\begin{figure}
    \centering
    \tikzset{grid_lines/.style={very thin, color=blue!50}}
    \begin{tikzpicture}[scale=1]
        % trailing
        \draw (0,0) -- (0,1) -- (3,1) -- (3,0) -- cycle;
        % leading
        \draw (4,0) -- (4,1) -- (7,1) -- (7,0) -- cycle;
        % broadening
        \filldraw[fill=gray!20!white] (7,0) -- (7, 1) -- (7.5,1) -- (7.5,0) -- cycle;
        
        % particles
        \filldraw[black] (3,0.5) circle (2pt);
        % \filldraw[fill=gray!90!white, draw=black] (4,0.5) circle (2pt);
        \filldraw[fill=gray!90!white, draw=black] (4.2,0.5) circle (2pt);
        \filldraw[fill=white,draw=black] (4,0.5) circle (2pt);
        
        % labels
        \draw[|<->|] (0,-0.25) -- (3,-0.25) node[below,midway] {$ L'_0 $};
        \draw[|<->|] (3,-0.25) -- (4,-0.25) node[below,midway] {$ h'_0 $};
        \draw[|<->|] (4,-0.25) -- (7.5,-0.25) node[below,midway] {$ L'_0 + \Delta x'(t'_{02}) $};
        \node at (1.5,0.5) {$ f'_2(t'_{02}, x', u') $};
        \node at (5.75,0.5) {$ f'_1(t'_{02}, x', u') $};
        \draw[->] (1.5,1.25) -- (2.5,1.25) node[above, midway] {$ \beta_2 $};
        \draw[->] (5.75,1.25) -- (6.75,1.25) node[above, midway] {$ \beta_1 $};
        \draw (0,-0.75) -- (0,1.5);
        \node[rotate=90] at (-.25,0.375) {Stellar surface};
    \end{tikzpicture}
    \caption{As in Figure~\ref{fig:g1g2_a}. The circles denote particles in the two clouds as discussed in the text.}
    \label{fig:g1g2_b}
\end{figure}
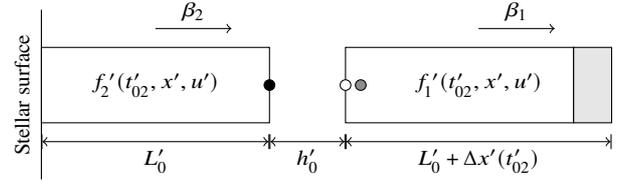

In case (1), we require particles with velocity $ \beta_2 $ at the leading edge of the trailing cloud, denoted by the black circle in Figure~\ref{fig:g1g2_b}, to catch up with particles with velocity $ \beta_1 $ in the trailing edge of the leading cloud, denoted by the gray circle, which are at a distance $ c\beta_1t'_{02} = (L'_0 + h'_0)(\beta_1/\beta'_{1-}) $ from the stellar surface so that their initial separation is $ (\beta_1/\beta'_{1-})(L'_0 + h'_0) - L'_0 $. The particles in the trailing cloud catch up to the particles in the leading cloud after a distance $ [(\beta_1/\beta'_{1-})(L'_0 + h'_0) - L'_0]/(\beta_2/\beta_1 - 1) $ is travelled by the latter particles. We may approximate this distance as $ 2\gamma_1^2[h'_0 + (L'_0 + h'_0)(1 - \gamma'^2_{1-}/\gamma_1^2)/2\gamma'^2_{1-}] \gtrsim 2\gamma_1^2h'_0 $, where we use $ \gamma_2 \gg \gamma_1 \gg \gamma'_{1-} $. It is clear that the distance required is greater than $ 300 R_* $, the upper limit for emission height \citep{Mitra17}, for $ \gamma_1 \gtrsim 1.2\times10^2 $. The minimum distance required is $ 200 R_* $ which is much larger than the emission height of most pulsars. Cloud overlap is therefore not possible in this case.

In case (2), we require particles with velocity $ \beta'_{2+} $ in the trailing cloud, denoted by the black circle in Figure~\ref{fig:g1g2_b}, to catch up with particles with velocity $ \beta'_{1-} $ in the leading cloud, denoted by the white circle. The trailing cloud catches up to the leading cloud after a distance $ h'_0/(\beta'_{1-}/\beta'_{2+} - 1) $ is travelled by the trailing particles. As in case (1), we may approximate this distance as $ \gtrsim 2\gamma'^2_{1-}h'_0 $, where the approximate form applies for $ \gamma'^2_{2+} \gg \gamma'^2_{1-} \gg 1 $. In this case cloud overlap is possible for most pulsars except for those where $ L_{\rm max} $ is small.

\subsubsection{Cloud overlap implications}

As discussed above, cloud overlap in the sense where the distributions lie on top of each other is not possible for two reasons: (1) relativistic streaming causes distributions to elongate (and fractionate) at different rates depending on the streaming Lorentz factor and energy spread; and (2) the distance required for overlap of bulk of the plasma far exceeds relevant emission heights in pulsars. Overlap in the sense that the slowest particles in a leading cloud is overtaken by the fastest particles in a trailing cloud is possible -- but not for all pulsars and pulsar parameters. Overlap in this sense relies on elongation of distributions due to relativistic effects.

There are two difficulties that arise. First, in the tail of the leading cloud, plasma number density can be much smaller than the number density of the rest of the cloud. The nature of the plasma wave supported by the cloud is then altered in the tail -- at the very least the plasma frequency is changed significantly. Second, it is most likely that the two clouds have similar number densities in their respective rest frames. As such, the leading edge of the trailing cloud will have a much larger number density than the tail of the leading cloud. Even if a plasma wave is supported by the tail of the leading cloud, the dispersion behaviour of the plasma will be dominated by the trailing cloud once it catches up. Therefore, the leading cloud cannot be considered the `background' plasma and responsible for dispersion properties with the trailing cloud acting as a `beam' and perturbing the dispersion properties of the background.

The second difficulty appears to be resolved if one considers the trailing cloud to be the background with the tail of the leading cloud acting as a beam. In this scenario, the portion of the tail of the leading cloud that is overlapping with the trailing cloud is very narrow in energy. This implies that it may be possible, in principle, for wave growth to occur through reactive instability. The bulk of the plasma in the pulsar frame is to propagate with streaming Lorentz factor $ \sim 10^2 {\rm -} 10^3 $. The requirement that the two clouds are separated in energy (RMMb,c) would necessitate very slow moving leading clouds. The possibility of such slowing moving clouds as well as actual possibility of reactive instability need further investigation. Kinetic instability is not viable for plausible pulsar parameters.

\end{document}